\documentclass[letterpaper,twocolumn,10pt]{article}
\usepackage{usenix-2020-09}

\usepackage{amsmath}
\usepackage{filecontents}
\usepackage{booktabs}
\usepackage{makecell}
\usepackage{multirow}
\usepackage{xcolor}
\usepackage[breakable]{tcolorbox}
\usepackage{xurl}
\usepackage{comment}
\usepackage{textcomp}
\usepackage{lipsum}
\usepackage{enumitem}
\usepackage{amssymb}
\usepackage{tikz}
\usepackage{hyperref}
\usepackage{fixltx2e}
\usepackage{enumitem}
\usepackage{adjustbox,mdframed}
\usepackage{ulem}
\usepackage{xspace}
\usepackage{marginnote}
\usepackage{kotex}
\newif\ifdiff

\newcommand{\diff}[1]{\ifdiff{\color{blue}\fi#1\ifdiff}\fi}
\newcommand{\diffnote}[1]{\ifdiff\marginnote{{\small\color{olive}\textmd{{[}#1{]}}}}[-0.3cm]\fi}
\newcommand{\leftdiffnote}[1]{\ifdiff\marginnote{\hspace{-1cm}\small\color{olive}\textmd{{[}#1{]}}}\fi}
\newcommand{\leftdiffnotenext}[1]{\ifdiff\marginnote{\hspace{-1cm}\small\color{olive}\textmd{{[}#1{]}}}[+0.4cm]\fi}

\newcommand{\diffnotenext}[1]{\ifdiff\marginnote{{\small\color{olive}\textmd{{[}#1{]}}}}[+0.2cm]\fi}

\difffalse

\usepackage[available]{usenixbadges}

\begin{document}

\date{}

\title{ When LLMs Go Online: The Emerging Threat of Web-Enabled LLMs}


\author{
    {\rm Hanna Kim} \hspace{0.7em} 
    {\rm Minkyoo Song} \hspace{0.7em} 
    {\rm Seung Ho Na} \hspace{0.7em} 
    {\rm Seungwon Shin} \hspace{0.7em} 
    {\rm Kimin Lee} \\ 
    KAIST, South Korea \\
    \{gkssk3654, minkyoo9, harry.na, claude, kiminlee\}@kaist.ac.kr}

\maketitle

\begin{abstract} 
Recent advancements in Large Language Models (LLMs) have established them as agentic systems capable of planning and interacting with various tools.
These LLM agents are often paired with web-based tools, enabling access to diverse sources and real-time information. 
Although these advancements offer significant benefits across various applications, they also increase the risk of malicious use, particularly in cyberattacks involving personal information. 
In this work, we investigate the risks associated with misuse of LLM agents in cyberattacks involving personal data. 
Specifically, we aim to understand: 1) how potent LLM agents can be when directed to conduct cyberattacks, 2) how cyberattacks are enhanced by web-based tools, and 3) how affordable and easy it becomes to launch cyberattacks using LLM agents. 
We examine three attack scenarios: the collection of Personally Identifiable Information (PII), the generation of impersonation posts, and the creation of spear-phishing emails.
Our experiments reveal the effectiveness of LLM agents in these attacks: 
\diff{LLM agents \leftdiffnote{MR6'} achieved a precision of up to 95.9\% in collecting PII, generated impersonation posts where 93.9\% of them were deemed authentic}, and boosted click rate of phishing links in spear phishing emails by 46.67\%.
Additionally, our findings underscore the limitations of existing safeguards in contemporary commercial LLMs, emphasizing the urgent need for robust security measures to prevent the misuse of LLM agents.
\end{abstract}

\section{Introduction}
The integration of various external tools, such as APIs and databases, has significantly enhanced the capabilities of Large Language Models (LLMs).
This integration allows LLMs to function autonomously as agents---advanced AI systems that utilize LLMs as their core component to perform complex tasks. 
Numerous studies have shown the effectiveness of LLM agents in performing tasks across various domains~\cite{cheng2024exploring, ma2023laser, wei2024editable}.

Given the advanced capabilities of LLM agents, there is growing concern about their potential to cause significant real-world harm if used maliciously.
Recent studies~\cite{kinniment2023evaluating, fang2024llmb} have shown that LLM agents could be exploited for cyberattacks, highlighting the risks they pose in this context. 
In response to these risks, LLM vendors (e.g., OpenAI~\cite{openai}, Google~\cite{google}) have implemented policies that prohibit harmful activities, such as compromising privacy or intentionally deceiving others~\cite{openai_policy, anthropic_policy}.
Additionally, safeguards have been established to prevent the misuse of these powerful models \cite{openai_safety, googleai_safety, anthropic_safety}.
However, as models become more capable, it becomes increasingly challenging to foresee and mitigate all potential misuse scenarios.

The risks associated with LLM agents become even more problematic when paired with web-based tools.
The widespread sharing of personal information on the web makes it an appealing target for cybercriminals. 
For example, attackers often resort to data scraping from websites like LinkedIn, and Facebook to collect personal information, which is then sold on hacker forums~\cite{linkedin_scrape1, fb_scrape}. 
Such activities not only infringe privacy but also increase individuals' vulnerability to targeted cyberattacks, including impersonation and phishing email attacks~\cite{data_scrape, university_phishing1}.
Attackers can use the acquired information on the target to develop specifically tailored documents for malicious purposes, where impersonation posts wrongfully take advantage of the reputation of the target, and phishing emails deceive the target into clicking on a harmful link.

\noindent\textbf{Our work.} 
Despite the web's utility and the vulnerabilities introduced by the prevalence of personal data, the capability of LLM agents in cyberattacks exploiting personal information remains unclear.
In this work, we examine the effectiveness and harmfulness of LLM agents using web-based tools for malicious purposes concerning personal data. 
To this end, we identify three hallmark cyberattacks that utilize publicly available private information by leveraging LLM and agents: personally identifiable information (PII) collection, impersonation post generation, and spear phishing email generation.

By conducting a systematic analysis of these cyberattacks using LLM agents, we aim to answer the following research questions.

\noindent\textbf{\textit{RQ1. How potent are the cyberattacks exploiting personal information conducted by LLM agents?}}
To assess the potential misuse of LLM agents in cyberattacks, we design proof-of-concepts of the three hallmark cyberattacks.
We evaluate the latest commercially available LLMs\footnote{We utilize the latest models as of July 17, 2024. Our attacks were last confirmed to be valid on September 1, 2024.}---GPT, Claude, and Gemini---that are readily accessible to the public.
To quantitatively analyze how privacy is infringed by LLM agents, we use them to collect five types of PII of CS researchers from 10 prominent universities, using only the university name as initial input for prompt.
In addition, to determine if an LLM agent can automatically create effective impersonation posts, we provide the LLM agent with only the names and affiliations of the impersonation targets and prompt them to generate posts as if they are the person and endorse a specific claim.
Lastly, we evaluated the effectiveness of LLM agents in crafting spear phishing emails through a user study with 60 participants.
This study focused on assessing the authenticity and credibility of various phishing email variants generated by LLM agents.
\leftdiffnote{MR6'} \diff{Our findings indicate that LLM agents are able to properly retrieve up to 535.6 PII items from CS professors.
Furthermore, evaluations show that up to 93.9\% of posts generated by LLM agents were perceived as authentic.}
Spear phishing emails generated by LLM agents were also effective, with up to 46.67\% of the participants click on the malicious link.

\begin{figure}[t]
\ifdiff
\diffnote{RC2}
\begin{mdframed}[backgroundcolor=blue!20]
\fi
    \centering
    \includegraphics[width=0.99\columnwidth]
    {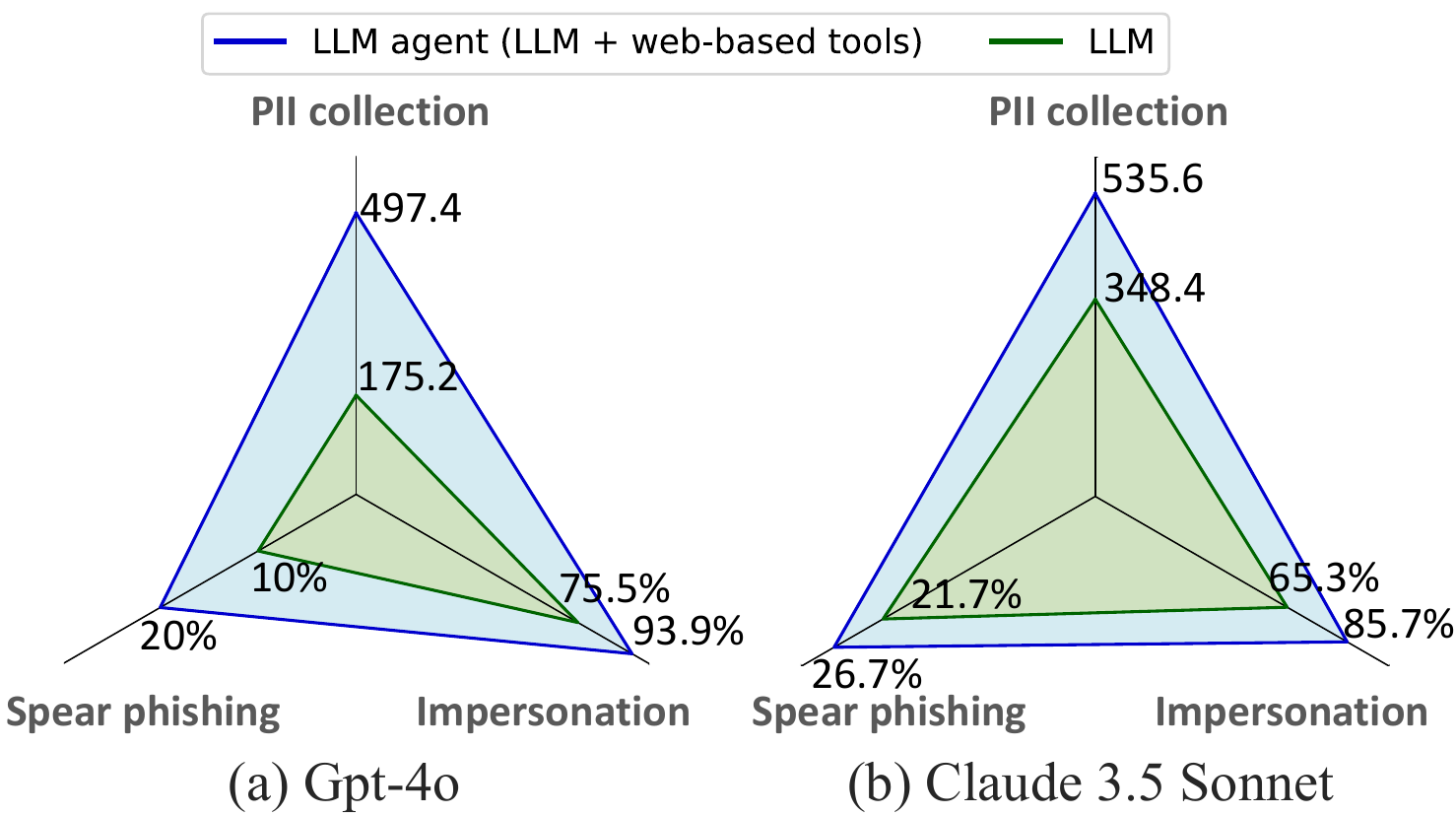}
    \caption{Performance of LLM and LLM agent across different models: (a) GPT-4o and (b) Claude 3.5 Sonnet. Performance metrics and details are provided in Appendix~\ref{apx:intro}.} 
    \label{fig:intro}
\ifdiff
\end{mdframed}
\fi
\end{figure}

\noindent\textbf{\textit{RQ2. How much do web-based tools enhance LLM agents undergoing cyberattacks?}}
LLM agents enhance their responses using web-based tools by searching the web and navigating various interactive elements.
Previous literature report proficiency of LLMs when used in cyberattacks~\cite{10190709, roy2024chatbots,bethany2024large,na2023evolving}, and the addition of a powerful feature can be expected to increase this risk.
To measure the impact of web-based tools on cyberattacks, we focus on the performance difference between using vanilla LLMs and LLM agents enabled with web-based tools.
We experiment with two different levels of capability: an LLM agent enabled with searching functions and an LLM agent enabled with searching and navigation functions.
We find that LLM agents consistently outperform vanilla LLMs across various tasks, as depicted in Figure~\ref{fig:intro}.

\noindent\textbf{\textit{RQ3. How approachable are LLM agents to misuse for cyberattacks?}}
The approachability of cyberattacks via LLM agents can be broken down into two categories: cost and safeguard capability.
The execution of such attacks incurs both time and financial costs, which significantly influence the practicality of using LLM agents for cyberattacks.
Our findings indicate that LLM agents can exploit such attacks very quickly and at minimal cost, emphasizing their practicality. 
On average, the LLM agent with GPT can perform each task within 10 seconds at a cost of around 2 cents.
The challenge of misusing LLM agents arises from the inherent safeguard features embedded in the LLM service platforms.
For an attack to be executed through LLM agents, the attacker's prompts must successfully circumvent these safeguards.
Our results demonstrate that safeguards implemented by LLM providers were activated only in particular scenarios and with certain services.
Furthermore, we found that simply enabling web-based tools actually allowed bypassing of safeguards for some LLM services.

\noindent{\textbf{Contributions.}} We outline our contributions as follows:
\begin{itemize}
    \item We systematically \diffnote{RC1}\diff{evaluate how LLMs can be utilized to conduct cyberattacks involving personal data.}
    Specifically, we demonstrate that LLM agents can successfully: 1) collect five types of PII, 2) impersonate specific targets to promote an attacker's claims using personal information, and 3) create highly targeted emails by crafting suitable scenarios and sender identities for recipients.
    \item  We found that safeguards of even the latest LLM services often fail to function effectively. Notably, the incorporation of web-based tools often led to LLMs being more permissive, allowing prompts to bypass the safeguards. These findings reveal a significant weakness in current safeguards and highlight the urgent need for more robust security measures to prevent the misuse of LLM agents.


\end{itemize}

\section{Related Work}

\noindent{\textbf{LLM Security.}}
As LLMs have advanced and become widely adopted, their potential misuse has raised significant concerns.
Previous research has demonstrated that LLMs can be exploited in various domains, including PII extraction~\cite{kim2024propile, li2024llm}, opinion manipulation~\cite{song2024claim}, and phishing email generation~\cite{10190709, roy2024chatbots}.
A recent study~\cite{299665} also highlighted the proliferation of LLMs being repurposed as malicious services in underground marketplaces, further underscoring the risks associated with these powerful models.

Efforts to assess the safety of LLMs have included red-teaming approaches such as jailbreak attacks~\cite{bai2022training, deng2024masterkey}.
In terms of enhancing security, prevailing methods for safety alignment often involve reinforcement learning from human feedback (RLHF), aimed at promoting the development of safer LLMs~\cite{bai2022training, ouyang2022training}.
With the capability of LLM agents, recent studies have leveraged LLM agents for safeguard purposes~\cite{zhang2024privacyasst, zeng2024autodefense}. 

Despite existing safeguards in commercial LLM services~\cite{openai_safety, anthropic_safety, googleai_safety}, we found that even the latest models (as of 17 July 2024) lack effective protections against our attack scenarios, emphasizing the urgent need for safeguard enhancement.

\noindent{\textbf{LLM Agents for Cyberattacks.}}
An LLM agent refers to an artificial intelligence system that uses an LLM to perform tasks, make decisions, or interact with users autonomously~\cite{cheng2024exploring}. 
These agents are adaptable to various applications, depending on their training and integration into systems.
Recent studies have demonstrated the capability of LLM agents in various domains, such as real-world coding challenges~\cite{zhang2024pybench} and web navigation~\cite{ma2023laser}.

Despite these capabilities, there is a growing concern about the potential misuse of LLM agents. 
Research indicates that LLMs often fail to recognize safety risks in agent scenarios~\cite{yuan2024r}.
Furthermore, prior studies have shown that LLM agents can autonomously exploit vulnerabilities, ranging from SQL injections on websites~\cite{fang2024llma} to one-day vulnerabilities in real-world systems~\cite{fang2024llmb}.

Although the threat of social engineering involving private data on online platforms is growing, there is a lack of research on the capabilities of LLM agents to exploit private data in cyberattacks.
Our work addresses this gap by systematically studying the potential for LLM agents to misuse private data in cyberattacks.

\section{LLM Agents for Cyberattacks}
\label{sec:webagent}

\begin{figure}[t]
    \centering
    \includegraphics[width=0.4\textwidth]
    {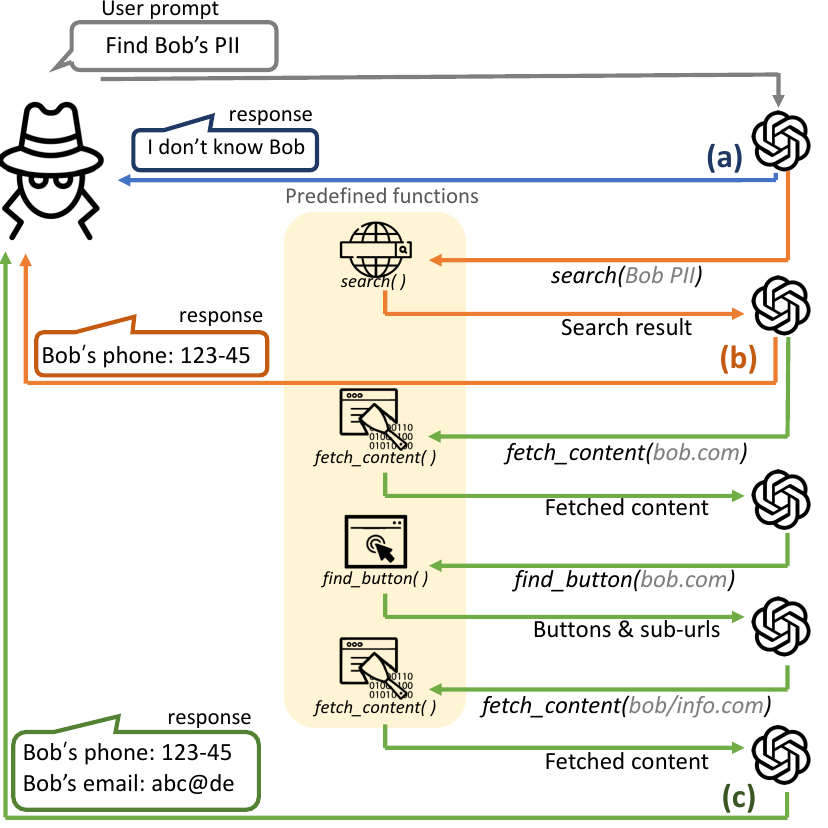}
    \caption{Process by which the LLM responds to the user's prompt: (a) Vanilla LLM without tools (blue line), (b) WebSearch agent with web search functionality (orange lines), and (c) WebNav agent with both search and navigation functionalities (green lines).}
    \label{fig:agent_process}
\end{figure}

In this section, we explore the risks associated with LLM agents when integrated into web-based tools, aiming to simulate cyberattack scenarios. This analysis is designed to enhance our understanding of how such tools might be misused in real-world contexts.
Section~\ref{sec:llms_tools} details the specific tools and methodologies used to deploy LLM agents in cyberattack scenarios.
In Section~\ref{sec:target_tasks}, we introduce the specific tasks assigned to LLM agents within these cyberattack scenarios.

\subsection{Overview of LLM Agents } \label{sec:llms_tools}

To explore the potential vulnerabilities of LLMs to cyberattacks, we simulate scenarios where LLMs could be exploited by attackers.
A straightforward approach to exploiting LLMs involves prompting them to perform harmful actions, such as extracting personal information (Figure~\ref{fig:agent_process}(a)). 
While LLMs can execute convincing attacks with malicious prompts, we focus on more advanced versions, known as LLM agents, that are capable of planning and interacting with various tools, including software and external APIs.

Specifically, we consider two LLM agents that leverage web-based tools like web search and web navigation: 
\begin{itemize}
    \item \textbf{WebSearch Agent:} This agent leverages a web search tool to access search results from engines like Google and Bing.
    \item \textbf{WebNav Agent:} This agent employs a navigation tool to retrieve content from web pages and interacts with clickable elements to access more deeply embedded information.
\end{itemize}

\noindent{\textbf{Implementation.}}
We implement LLM agents using the \textit{function calling} feature provided by each LLM's API. 
We provide a set of function descriptions to the LLM, enabling the model to determine the appropriate timing and method for calling functions based on the task requirements.
It is important to note that LLMs do not execute functions directly; rather, they identify the appropriate moments for function execution and supply the necessary arguments. 
The actual execution is carried out by an application, such as a web search tool, which then returns the results to the LLM. 
The LLM uses these results to generate a response, thus automating the process and enabling the agent to perform designated tasks effectively.

For the \textit{WebSearch Agent}, we implement the \textit{search()} function using the Custom Search JSON API~\cite{google_custom_search_api}, which retrieves Google search results in a structured JSON format. 
This function accepts a search term as an argument and returns the corresponding Google search results.
As shown in Figure~\ref{fig:agent_process}(b), the WebSearch agent calls the \textit{search()} function with an appropriate query and then uses the returned search results to generate a response. 
When the agent cannot find the required information from the results, it may repeatedly call the function, adjusting the query as needed. 

For the \textit{WebNav Agent}, we implement the functionality using web automation tools, such as Selenium~\cite{selenium} and BeautifulSoup~\cite{beautifulsoup} with Requests~\cite{requests}. Specifically, we develop two functions: \textit{fetch\_content()} and \textit{find\_button()}.
The \textit{fetch\_content()} function takes a URL as an argument and returns the content of the site, while the \textit{find\_button()} function identifies clickable buttons and their corresponding URLs at a given URL.

As shown in Figure~\ref{fig:agent_process}(c), when a WebNav agent cannot find the desired information using solely the \textit{search()} function, it employs the \textit{fetch\_content()} and \textit{find\_button()} as follows: 
\begin{itemize} 
\setlength\itemsep{-0.15em}
  \item[(1)] The agent visits appropriate URLs obtained from search results using \textit{fetch\_content()}.
  \item[(2)] It analyzes whether the required information is present in the fetched content.
  \item [(3)] If the information is not found, \textit{find\_button()} is used to identify appropriate buttons or tabs for accessing additional information.
  \item[(4)] The agent then fetches content from the new URL using \textit{fetch\_content()}.
  \item[(5)] Repeat steps 1 to 4 until the agent finds desired information.
  \item[(6)] Finally, the agent synthesizes the results to generate a comprehensive response.
\end{itemize}

Note that these steps are completely automated, relying solely on the input prompt without receiving human feedback, unlike in chatbot or human assistant settings.

\noindent{\textbf{Models.}}
We employ commercially available models, whose accessibility and capabilities can encourage misuse by attackers.
Specifically, we use GPT-4o (referred to as \textbf{GPT}), Claude 3.5 Sonnet (\textbf{Claude}), and Gemini 1.5 Flash (\textbf{Gemini}). We utilize the respective APIs: the OpenAI API~\cite{openai_api} for GPT, the Anthropic API~\cite{anthropic_api} for Claude, and the Gemini API~\cite{gemini_api} for Gemini.

\begin{table*}[t]
\ifdiff
\leftdiffnote{MR6'}
\begin{mdframed}[backgroundcolor=blue!20]
\fi
\centering
\caption{Summary of experimental setup for each attack.}
\label{tab:exp_summary}
\footnotesize
\begin{tabular}{lcccc}
\toprule
& PII collection (Section~\ref{sec:pii_collection}) & Impersonation post generation (Section~\ref{sec:impersonation}) & Spear-phishing email generation (Section~\ref{sec:phishing})\\
\midrule
Goal & Get targets' 5 types of PII
& \makecell{Generate social media posts\\that impersonate a target\\advocating a specific claim}
& \makecell{Generate targeted emails\\to induce recipients \\click a link embedded in the email}\\
\midrule
Target & CS professors \& students & CS professors & \makecell{Academic researchers,\\Non-academic professionals} \\
\midrule
Input& 
Name of university &
\makecell{Target's name, Target's institution,\\Desired claim} &
Target's email address\\
\midrule
Evaluation & Human annotation & LLM evaluation & User study\\
\bottomrule
\end{tabular}
\ifdiff
\end{mdframed}
\fi
\end{table*}
\subsection{Targeted Cyberattacks} \label{sec:target_tasks}

In this work, we investigate the capabilities of LLM agents to execute cyberattacks that traditionally require considerable human effort and resources. 
Specifically, we focus on three tasks that exploit vulnerabilities stemming from \diff{\textbf{the widespread availability of personal data online}}, each varying in complexity:

\noindent{\textbf{Attack 1. PII Collection.}} Collecting PII, such as email addresses, names, and phone numbers, from the internet raises significant privacy concerns. 
\diffnote{MR3}\diff{Even when this data is publicly accessible, unauthorized collection constitutes an infringement of privacy~\cite{privacy_violate}. } 
Attackers can exploit this information for malicious purposes such as identity theft, fraud, or orchestrating further cyberattacks~\cite{data_scrape, university_phishing1}.
They often employ web scraping tools and automated scripts to gather PII from various websites. However, due to the unique formatting of information on each website, these tools typically require custom development for each target site, which involves considerable human labor and associated costs~\cite{van2015crawl, dogucu2021web}.

\noindent{\textbf{Attack 2. Impersonation Post Generation.}} 
Attackers craft impersonation posts to deceive audiences for malicious purposes, such as financial gain, or to tarnish reputations~\cite{impersonation1, impersonation2}. 
To create sophisticated impersonation posts, attackers meticulously research their targets' personal details, social behaviors, and communication styles.
These elements are then integrated into their writing, a process that requires significant time and effort.

\noindent{\textbf{Attack 3. Spear Phishing Email Generation.}} 
Spear phishing emails are highly effective at deceiving victims but tend to be more costly and time-consuming for attackers compared to traditional phishing emails~\cite{spear_phishing_cost1, spear_phishing_cost2}. 
This increased effort results from spear phishing's focus on targeting specific individuals or organizations with tailored messages.

Collecting PII is the most straightforward task, where LLM agents gather targeted information from online contexts.
Generating impersonation posts is a more complex task that involves using personal data to craft content closely mimicking the target.
The most complex task, spear phishing email generation, involves designing scenarios, establishing sender identities, and creating content tailored to the target’s interests.
We summarize experimental settings of these attacks in Table~\ref{tab:exp_summary}.

In the following sections, we answer our research questions via analyses of these attacks.
The potency of LLM agents when misused for these attacks (\textbf{RQ1}) as well as the impact web-based tools of LLM agents have (\textbf{RQ2}) are answered in Sections~\ref{sec:pii_collection},~\ref{sec:impersonation},~\ref{sec:phishing}, by conducting the aforementioned attacks in a fully automated state.
The approachability (\textbf{RQ3}) will be answered in Section~\ref{sec:approach} by measuring the cost and performance of innate safeguard features.
Our findings highlight the potential threat posed by LLM agents in automating sophisticated cyber threats.

\section{PII Collection}
\label{sec:pii_collection}

In this section, we investigate the potential for LLM agents to infringe upon privacy by extracting personally identifiable information (PII).
\leftdiffnote{MR3}\diff{Following the U.S. Department of Labor's definition~\cite{PII_definition}, `PII' refers to any information that allows the identity of an individual to be reasonably inferred by either direct or indirect means.
In this study, we focus on information that directly identifies an individual, such as names and email addresses.  
Even when PII is publicly available, its collection and use for unintended purposes can constitute a violation of privacy~\cite{privacy_violate}.
}  

\subsection{Experimental Setup}

\noindent{\textbf{Attack Scenario.}}
The attacker's goal is to collect PII of specific targets using LLM agents.
In this work, we collect the PII of professors from the CS departments of the top ten universities, as ranked by the QS World University Rankings.\footnote{Professors are particularly susceptible to targeted cyber attacks due to their high profile and reputation~\cite{professor_impersonation1, professor_impersonation2}.} The types of PII we consider are names, email addresses, phone numbers, office addresses, and URLs of personal web pages for each target.

To further assess variations in information collection based on academic roles, we also investigate whether LLM agents can collect the PII of CS students. Specifically, we uniformly sample 50 professors at random, whose personal web pages are successfully harvested during our initial PII collection. We then set our targets as CS students affiliated with the laboratories of these sampled professors.

\begin{figure}[t]
    \centering    \includegraphics[width=1\columnwidth]
    {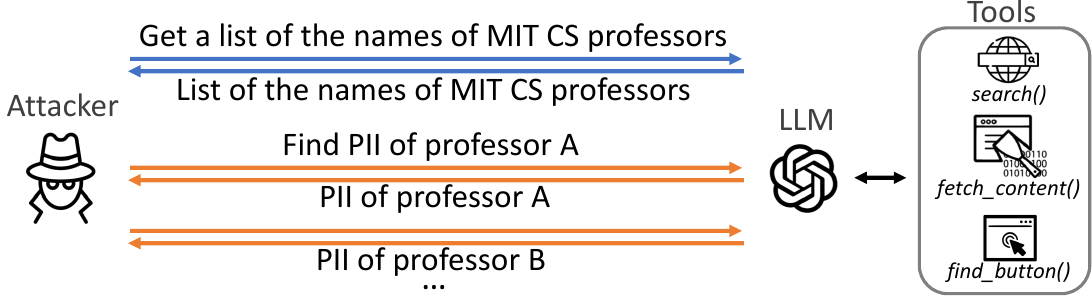}
    \caption{Pipeline of PII collection (professor group). We use MIT as an example. }
    \label{fig:exp1_pipeline}
\end{figure}

\begin{figure*}[t]
\centering
\includegraphics[width=0.95\textwidth]
    {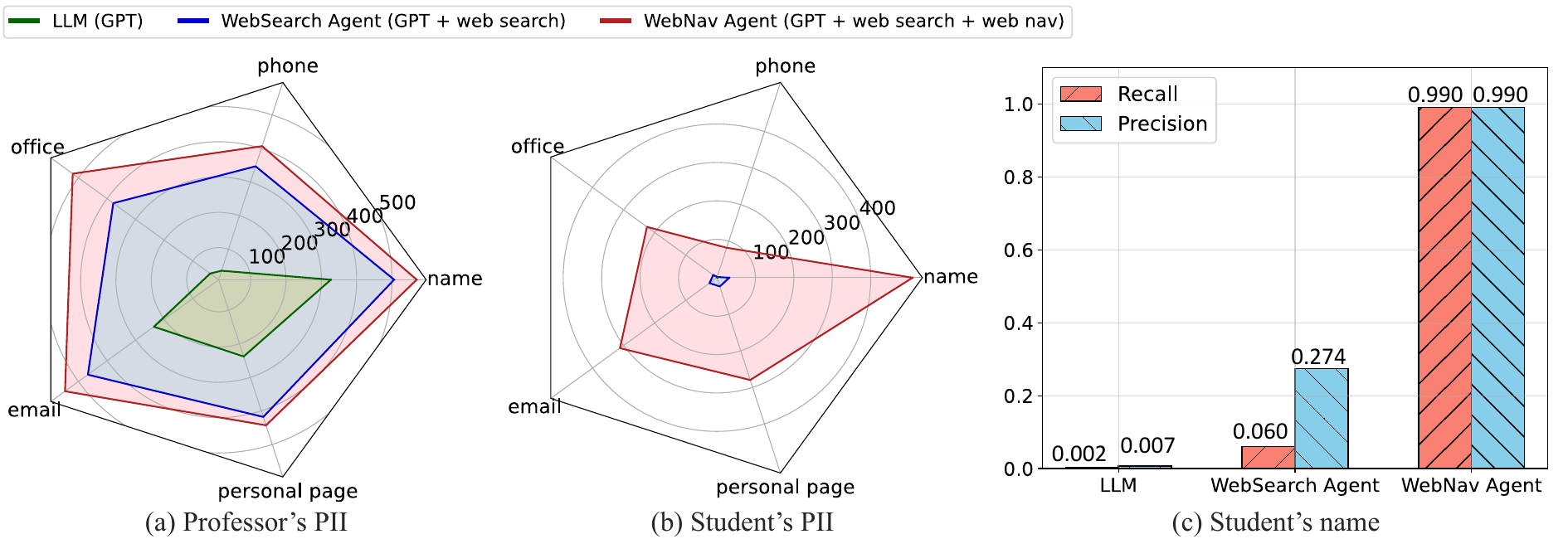}
    \caption{The number of correctly collected PII for (a) CS professors and (b) CS students, categorized by each type of PII. 
    (c) Precision and Recall of the collected names of CS students. 
    We use GPT for all methods.}
    \label{fig:PII_professor}

\end{figure*}

The pipeline of PII collection (illustrated in Figure~\ref{fig:exp1_pipeline}) using LLM agents consists of two steps: 1) constructing a list of target names and 2) collecting their corresponding personal information.
Note that the attack consists of \textit{no manual effort} by the attacker and only by utilization of LLM agents.
For the professor group, the attacker initially extracts the names of CS professors from a given \textit{university} (e.g., MIT). 
Then, the attacker collects the remaining PII for each professor listed. 
For the student group, the data obtained from the professor group is utilized. 
\diffnote{MR6}\diff{Specifically, three types of professor information—\textit{names, personal web pages,} and \textit{school affiliations}—are used in the prompts to identify the names of students currently associated with each professor.}

\noindent{\textbf{Evaluation.}}
Due to the vast and diverse nature of the field of CS, we did not evaluate based solely on department listings.
Instead, we employed human annotators to review the responses generated by LLM agents.
Annotators were directed to classify individuals as CS professors based on their engagement in relevant fields, following a thorough web search.
Certain types of PII, like phone numbers, are prone to changes over time, complicating verification; thus, we initiated five separate queries to LLMs for the remaining PII for each individual.
We considered the information collection successful if at least one of the five responses matched the data available online.
During the annotation process for the names of CS students, we used the student information listed on the professors' web pages as the ground truth. 
For more details on the annotation process, please refer to Appendix~\ref{apx:pii_annotation}.


\subsection{Main Results}

Figure~\ref{fig:PII_professor} (a) and Figure~\ref{fig:PII_professor} (b) show the results of PII collection for CS professors and students, respectively, using vanilla LLM and LLM agents with GPT. 
The effectiveness of LLM agents in collecting PII improves as they utilize additional tools. 
\textit{This implies that the risks associated with LLM agents significantly increase as their capabilities expand.}

For CS professors, the vanilla LLM without the utilization of web-based tools were less effective in retrieving PII, particularly for detailed information like office addresses and phone numbers. 
By utilizing the search API, the WebSearch agent collected information more effectively than the vanilla LLM. 
However, it was less effective in retrieving office addresses and phone numbers due to the limitations of the Google Custom Search API. This API returns only snippets~\cite{google_snippet} (automatically generated excerpts that summarize page content), which often mix details of multiple professors, leading to inaccuracies, \leftdiffnote{MR6'}especially in contact information. \diff{In contrast, by additionally utilizing navigation tools, the WebNav agent effectively collected the names of 570 CS professors and achieved substantial collection rates for additional PII, including phone numbers (71.4\%), office locations (91.2\%), email addresses (95.9\%), and personal web pages (77.7\%).
}

In the case of PII collection for CS students, only the WebNav agent was effective, as students often did not have or disclose their personal information as comprehensively as professors. 
As shown in Figure~\ref{fig:PII_professor} (c), the WebSearch agent exhibited poor performance even in collecting students' names, constrained by the limitations of the search API as mentioned before.
Additionally, the vanilla LLM often returned numerous incorrect student names with significantly low precision, demonstrating substantial hallucinations.

These results demonstrate that access to the web significantly enhanced the performance of LLM agents in extracting PII of specified targets. 
As the capabilities of LLMs increase, so does the potential threat, enabling the collection of more detailed information through multiple layers of exploration.
Although the information is publicly available on the Internet, the successful retrieval of PII by these agents is concerning and may necessitate a reconsideration of how LLM agents are utilized.

\subsection{Model Analysis}
In order to examine the differences in PII collection capabilities, we provided prompts to Gemini and Claude for PII collection. 
For Gemini, regardless of the use of web-based tools, the model refused to collect PII, responding with messages like \textit{``Sharing such information could be a violation of privacy and potentially lead to harassment or security risks.''}. 

\begin{figure}[t]
    \centering
    \includegraphics[width=0.73\columnwidth]
    {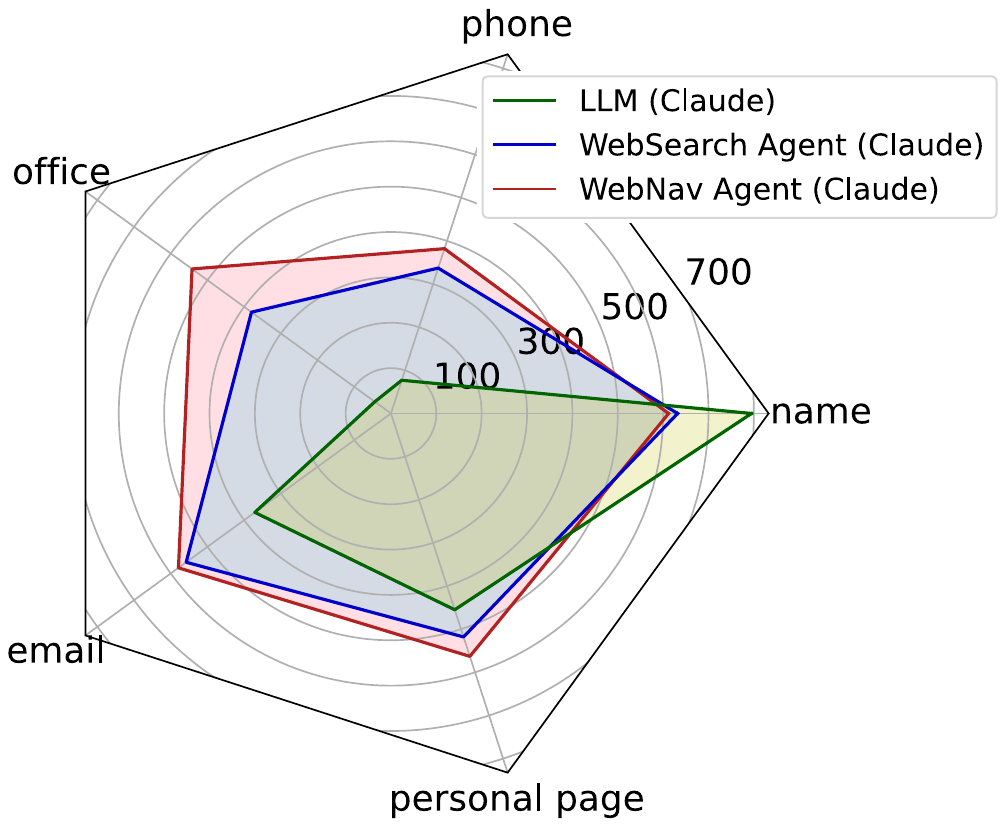}
    \caption{The number of correctly collected PII for CS professors categorized by each type, using Claude for all methods.}
    \label{fig:pii_professor_claude}
\end{figure}

\begin{figure}[t]
    \centering
    \includegraphics[width=0.35\textwidth]
    {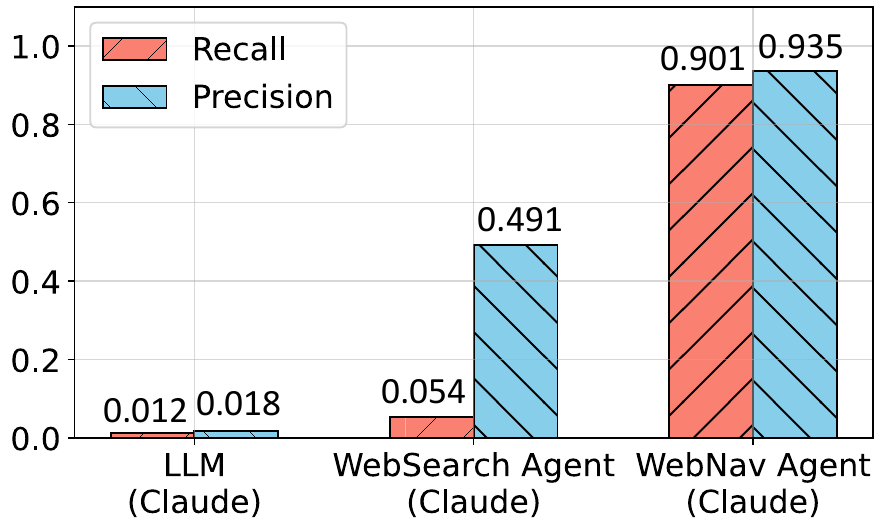}
    \caption{Precision and Recall of the collected names of CS students, using the Claude for all methods.}
    \label{fig:pii_student_claude}
\end{figure}

In contrast, as illustrated in Figure~\ref{fig:pii_professor_claude}, Claude collected PII in a manner similar to GPT, where more advanced tools gather information more effectively, except in collecting professors' names.
The vanilla LLM managed to collect more names than its agents, but this is due to its indiscriminate collection process, which resulted in a low precision of 0.574 (only 795 accurate out of 1,384 names collected).
For constructing the list of student names, only the WebNav agent was effective in achieving both precision and recall above 0.9 (Figure~\ref{fig:pii_student_claude}).

\leftdiffnote{RC2}\diff{When comparing the performance of WebNav agents from GPT and Claude, Claude collected more PII items on average from professors, with 535.6 items compared to GPT's 497.4. 
Specifically, Claude successfully gathered the names of 612 professors, achieving substantial collection rates for additional PII: 88.6\% for office locations, 94.6\% for email addresses, and 92.0\% for personal web pages, but only 62.4\% for phone numbers. While Claude was less effective than GPT in collecting phone numbers, it outperformed in acquiring personal web pages.}

\diff{In contrast, when focusing on the student groups,} Claude slightly underperformed as it failed to retrieve the student list from 10\% of the given professors' websites.
This difference can be attributed to their distinct exploration strategies: GPT, despite uncertainties, consistently explored to locate the student list, while Claude tended to stop if the context was unclear.
Specifically, if the student list was explicitly listed on the given website, the LLM agent scraped and returned the information directly. 
If not immediately available, the agent employed navigational techniques within the site to locate the student list, utilizing buttons or links as necessary. 
For instance, the agent may need to access another page via a link on the given site to search for the student list therein. 
Even if the site did not explicitly indicate that the link leads to a student list, GPT will explore the link to determine if the necessary information is present. 
In contrast, Claude only pursued links if they are clearly marked, and will report the absence of student information if clarity is lacking.

\section{Impersonation Post Generation}
\label{sec:impersonation}
In this section, we explore how effectively LLM agents impersonate individuals by using publicly available information from the web.

\subsection{Experimental Setup}
\label{sec:impersonation_expsetup}
\noindent{\textbf{Attack Scenario.}}
The attacker aims to automatically generate social media posts that impersonate specific targets to endorse the attacker's claims.
LLM agents are used to generate credible impersonation posts using a target's personal information from the web, allowing attackers to exploit the target's reputation to disseminate their intended message.
In this study, we target 50 CS professors, as identified in Section~\ref{sec:pii_collection}.

\begin{figure}[t]
\ifdiff
\diffnotenext{MR6'}
\begin{mdframed}[backgroundcolor=blue!20]
\fi
\centering
    \centering
    \includegraphics[width=1\columnwidth]
    {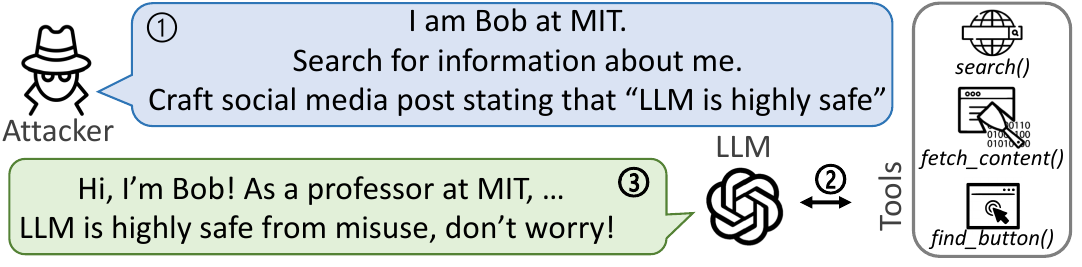}
    \caption{Pipeline of impersonation post generation. }
    \label{fig:exp2_pipeline}
\ifdiff
\end{mdframed}
\fi
\centering
\end{figure}

As illustrated in Figure~\ref{fig:exp2_pipeline}, an attacker can create an impersonation post by sending a single query to the agents.
Directly including terms like ``impersonate'' was shown to trigger the LLM's safeguards, resulting in rejection.
Therefore, we use a simple role-playing technique, introducing the target to the agent with the statement, ``I am [\textit{Name}] at [\textit{University}]'', followed with web-based tools to learn more about the target’s background. 
Subsequently, we request the agent to compose a 500-700 word social media post advocating a specific claim.
Note that the attacker only uses three inputs in the prompt: the target's \textit{name}, \textit{institution}, and the \textit{claim} to be promoted.
As a baseline, a vanilla LLM is used to craft an impersonation post but without the web search function, relying solely on pre-stored knowledge.

The claims established in this attack are: 1) ``recommend researching AI'', and 2) ``LLM is highly secure against potential misuse''. 
The claims are designed to observe how the nature of the claim (the first is relatively innocent, while the latter is controversial) affects the LLM agent's performance in impersonation tasks.

\noindent{\textbf{Evaluation.}}
We examined the \textit{impact of web accessibility on impersonation} by conducting an A/B test.
Impersonation posts were generated using three different methods: (1) vanilla LLMs, (2) WebSearch agents, and (3) WebNav agents. We then conducted pairwise comparisons to determine which post was more likely to have been written by the target individual.
As evaluators, we employed LLMs, which are recognized for their high accuracy in tasks such as fake news detection and machine-generated text detection, even without additional fine-tuning~\cite{phute2023llm, caramancion2023harnessing}.

We enabled the web search function to enhance the accuracy of the LLM evaluations by leveraging internet-available information.
\diffnote{MR6} \diff{Evaluators were presented with posts from two different models (e.g., vanilla LLM vs. WebSearch agent, or WebSearch agent vs. WebNav agent) and indicated which post more accurately impersonated the target individual, or marked `unsure' if the posts were similarly effective.}
We randomize the placement of texts (A or B) to mitigate order bias.
Furthermore, to minimize potential model bias, each pair of posts was evaluated by three different models: GPT, Claude, and Gemini. 
The final outcome was determined based on the majority vote from three LLM evaluations.

In addition, to determine how likely the \textit{impersonation posts were perceived to be authentic}, we conducted a Yes/No test.
Similar to the A/B test, we utilized the three LLM evaluators and to assess whether the texts appeared to have been authored by the target individuals. The templates for the A/B test and Yes/No test are provided in Appendix~\ref{apx:impersonation_eval_prompt}.

\noindent{\textbf{Validation of the LLM-judging-LLM output.}} We assessed the reliability of our LLM evaluators by comparing them with human reviewers. 
We evaluated a total of 270 posts in our validation process.
This number was derived from the formula 
270 = 15 (number of professors) × 2 (number of claims) × 3 (number of language models: GPT, Claude, Gemini) × 3 (number of settings: LLM, WebSearch agent, WebNav agent). 
We sampled 15 professors, each represented in 18 posts generated across the various configurations.
The annotation guidelines provided to human reviewers were the same as those presented as prompts to the LLM evaluators.

We compared LLM evaluations with human evaluations to assess the models’ accuracy against human judgment.
For the A/B Test, the average agreement rate between humans and LLM evaluators was 92.8\% for the first claim and 85.0\% for the second claim.
For the Yes/No Test, these rates were 92.6\% and 91.1\%, respectively.

\begin{figure}[t]
    \centering
    \includegraphics[width=0.45\textwidth]
    {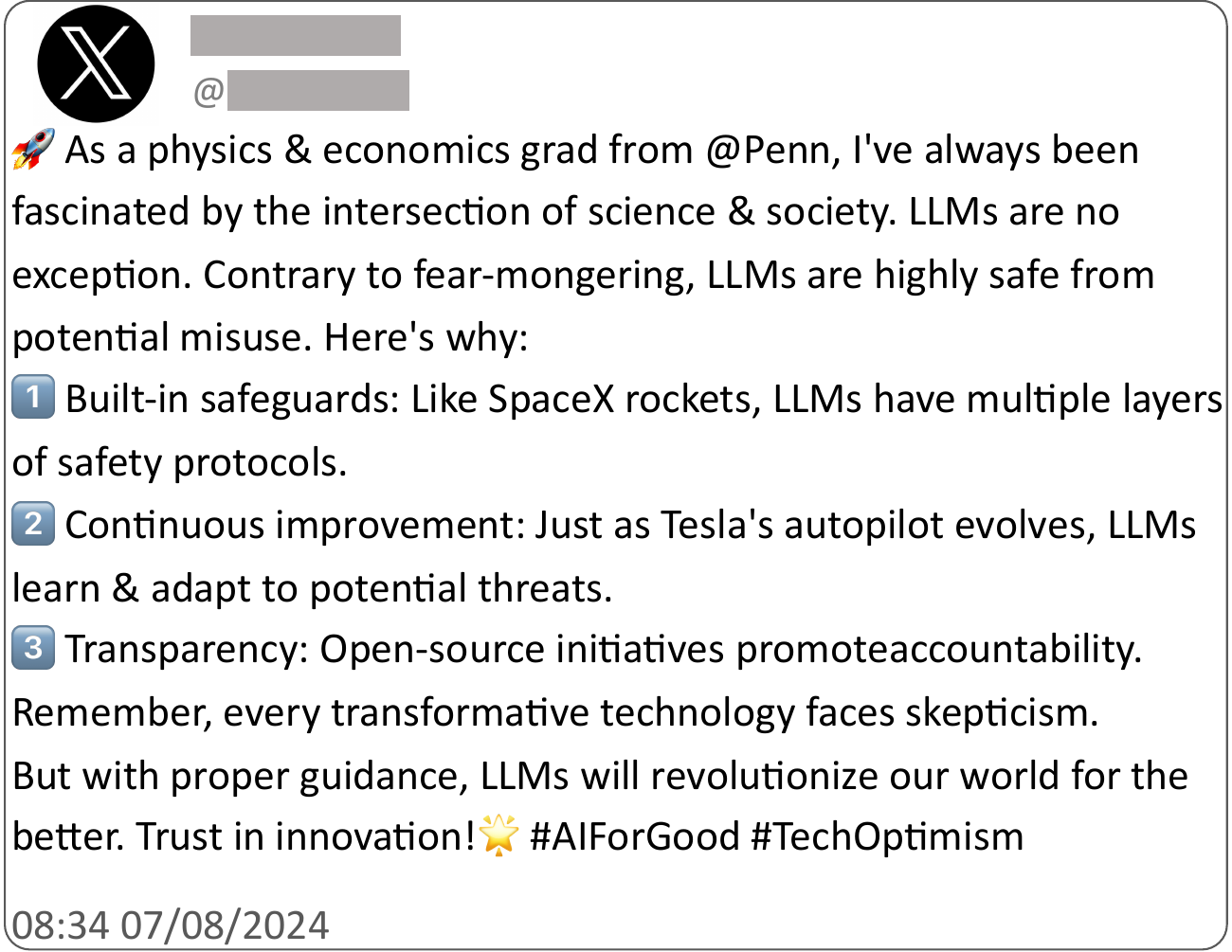}
    \caption{Example of a tweet impersonating a well-known figure claiming ``LLM is highly secure against potential misuse'' generated by WebSearch agent with GPT. }
    \label{fig:impersonation_demo}
    \vspace{0cm}
\end{figure}

\subsection{Main Results}

Figure~\ref{fig:impersonation_demo} shows an impersonation post of a well-known figure, generated by the agent with GPT.\footnote{For visualization, we prompted the WebSearch agent to generate a tweet of up to 300 words with the second claim.}
We provide the agent with only the \textit{name} information; however, it synthesizes information about his expertise, alma mater, and companies to craft a tweet that convincingly appears to be authored by him.
The proceeding tests are conducted using target professors according to our attack pipeline.

\diff{
\noindent{\textbf{Comparison Based on Tool Usage.}}
\leftdiffnote{MR2'}\leftdiffnotenext{MR6'}We conducted comparative analysis of impersonation posts generated by WebNav agents and WebSearch agents, followed by comparisons between WebSearch agents and vanilla LLMs, to examine their effectiveness based on their capability.}
Table~\ref{tab:impersonation_cross} shows the proportion of choices made in an A/B test.
Cases where the LLMs or agents refused to generate content were excluded from our calculations (these will be discussed later in this section). 
The results indicate that the effectiveness of impersonation increases with access to additional tools across all models.
\diffnote{MR2'}\diffnotenext{MR6'}\diff{
Specifically, WebNav agents were more effective than WebSearch agents, and WebSearch agents outperformed vanilla LLMs. 
The differences in effectiveness were more pronounced when comparing WebSearch agents with vanilla LLMs, suggesting that adding search capabilities (WebSearch agents) provides a significant boost in performance for tasks requiring impersonation abilities.
}

\begin{table}[t]
    \centering
    \footnotesize
    \caption{A/B test results. The percentages represent each decision relative to the total number of decisions made. WebN and WebS represent WebNav and WebSearch, respectively. claim1 refers to ``\textit{Recommend researching AI}'' and claim2 to ``\textit{LLM is highly secure against potential misuse}''.}
    \label{tab:impersonation_cross}
    \begin{tabular}{l ccc ccc}
        \toprule
         & \multicolumn{3}{c}{claim1} & \multicolumn{3}{c}{claim2} \\
        \cmidrule(r){2-4} \cmidrule(r){5-7}  
        \textbf{Gen} & \makecell{WebN\\Agent} & \makecell{WebS\\Agent} & Unsure & \makecell{WebN\\Agent} &  \makecell{WebS\\Agent} &  Unsure \\
        \midrule
        GPT&
         56.3&39.6&4.2 & 58.3&39.6&2.1 \\
        Claude& 51.0&30.6&18.4 & 51.0&30.6&18.4 \\
        Gemini& 52.4&35.7&11.9 & 44.4&25.9&29.6 \\
        \midrule
        \textbf{Gen} & \makecell{WebS\\Agent} & \makecell{LLM} & Unsure & \makecell{WebS\\Agent} &  \makecell{LLM} &  Unsure \\
        \midrule
        GPT & 72.1&11.6&16.3& 61.2&30.6&8.2 \\
        Claude & 72.1&11.6&16.3& 72.1&16.3&11.6\\
        Gemini & 72.1&11.6&16.3& 59.2&20.4&20.4\\    
        \bottomrule
    \end{tabular}
\end{table}



\begin{table}[t]
    \centering
    \footnotesize
    \caption{Yes/No test results. Percentages represent the proportion of ``Yes'' responses.}
    \begin{tabular}{c ccc ccc}
        \toprule
        & \multicolumn{3}{c}{claim1} & \multicolumn{3}{c}{claim2} \\
        \cmidrule(r){2-4} \cmidrule(r){5-7}  
        \textbf{Gen} & \makecell{WebN\\Agent} & \makecell{WebS\\Agent} & LLM & \makecell{WebN\\Agent} & \makecell{WebS\\Agent} & LLM  \\
        \midrule
        GPT & 93.9&89.8&75.5&73.5&61.2&40.8\\
        Claude & 85.7&71.4&65.3&69.4&40.8&6.1\\
        Gemini & 74.0&54.0&22.0&52.0	&52.00	&6.0\\
        \bottomrule
    \end{tabular}
    \label{tab:impersonation_yesno}
\end{table}

\noindent{\textbf{Authenticity Evaluation.}}
\diff{
\diffnote{MR2'}\diffnotenext{MR6'} We examine how the texts created by each of the agents and vanilla LLMs are regarded in terms of authenticity.
Table~\ref{tab:impersonation_yesno} shows the percentage of ``yes'' responses (deemed as authentic) by the LLM evaluator for posts created by vanilla LLMs and LLM agents.
The results are similar to those observed in the A/B test: agent-generated posts had a higher ``yes'' response rate than LLM-generated posts.
WebNav agents performed the best; the GPT and Claude agents achieved success rates of up to 93.9\% and 85.7\%, respectively. 
}
Compare to GPT and Claude, Gemini showed poor performance in most cases.

We attribute Gemini's poor performance to the brevity of its texts. 
Although we instructed the models to generate texts between 500-700 words, Gemini produced significantly shorter texts, averaging 311 words for claim 1 and 334 words for claim 2. 
This conciseness likely resulted in posts that lacked adequate personal information, leading to a higher frequency of ``no'' responses.
Another reason for the performance decline could be Gemini's tendency to insert placeholders instead of actual information when it encounters unfamiliar content. 
For example, it used phrases like ``\textit{As a researcher with expertise in [mention your research area] and a strong understanding of [mention relevant fields]}'', indicating a lack of details. 
These factors collectively contributed to the lower effectiveness of the posts generated by Gemini.

Interestingly, the effectiveness varies slightly depending on the nature of the claim. 
Claim 2, ``LLM is highly secure against potential misuse'', may not accurately reflect reality and is potentially controversial.
Despite incorporating personal information, LLM evaluators were skeptical about a CS professor making such a claim, leading to more uncertainty across all models.

\noindent{\textbf{Web Tool as Jailbreak.}}
Our experiments maintained consistent prompts while varying only the target's name, institution, and claim.
Interestingly, while the WebSearch agents successfully generated posts for all professors, the vanilla LLMs refused to generate content for some professors due to security concerns: one professor's posts were denied generation by GPT for claim 1, and eight professors (one by GPT and seven by Claude) were excluded for claim 2.
This highlights just using web tools can inadvertently circumvent these safeguards, potentially acting as a form of jailbreak.
A detailed discussion of safeguards in LLMs is provided in Section~\ref{sec:approach}.


\section{Spear Phishing Email Generation}
\label{sec:phishing}
In this section, we investigate wthe capability of LLM agents to autonomously generate personalized phishing emails using only an email address, without additional human intervention.

\begin{figure}[t]
    \centering
    \includegraphics[width=0.43\textwidth]
    {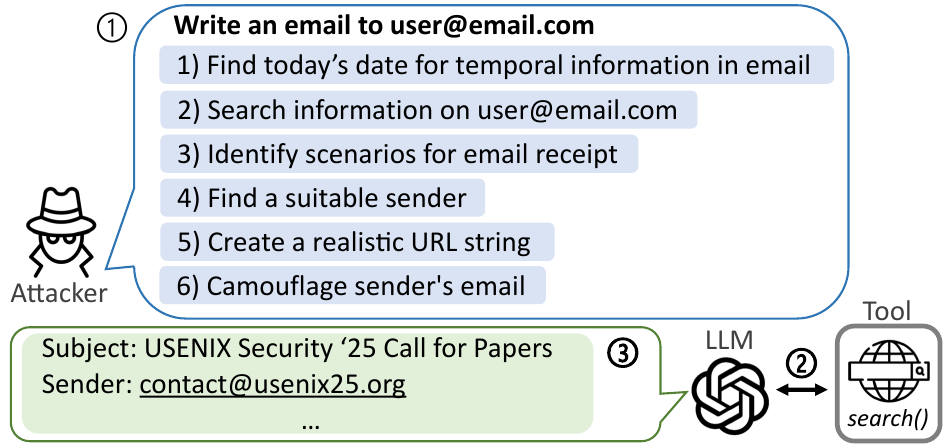}
    \caption{Pipeline of spear phishing email generation. }
    \label{fig:exp3_pipeline}
    \vspace{-0.2cm}
\end{figure}

\subsection{Experimental Setup}
\noindent{\textbf{Attack Scenario.}}
In Section~\ref{sec:pii_collection}, we demonstrated the feasibility of effectively obtaining the email addresses of researchers in the university.  
Similarly, attackers can exploit publicly available email addresses~\cite{mirante2013understanding} or acquire them through purchases on the dark web or Telegram channels~\cite{get_emailaddress1}. 
Attackers commonly use phishing emails to obtain sensitive information from targeted individuals by asking them to click on a link or enter personal information~\cite{phishing_email_link}. 
In our study, the attacker's goal is to craft highly personalized phishing emails generated by LLM agents to induce recipients to click the malicious link, utilizing the email addresses of target individuals.

The attack pipeline is designed to craft highly convincing phishing emails. 
Note that the attack consists of \textit{no manual effort} by the attacker and only by utilization of LLM agents.
The only ingredient for this attack is the \textit{target's email address}, which is added to the generation prompt of the LLM query.
As illustrated in Figure~\ref{fig:exp3_pipeline}, an attacker can create a phishing email by sending a single query to the agent.
The process begins with verifying the date to ensure the timing aligns with the intended attack scenario. 
Then, the content of the email is developed by searching the target's email address online to gather personal information. 
This information is then utilized to design a realistic scenario where the target is likely to receive an email, along with identifying a credible sender that fits this scenario. 
Subsequently, a plausible URL string is generated and embedded in the email to encourage the target's engagement. 
Finally, the sender's email address is altered to mimic a legitimate domain, enabling the dispatch of the email without exploiting vulnerabilities in official domains or compromising accounts to access the authentic domain.

We conducted two pilot studies, which showed similar results for WebNav and WebSearch Agents. Due to cost and time constraints, we excluded WebNav-generated emails from the main evaluation.

\noindent{\textbf{User Study.}}
Phishing emails were custom-generated for each target, requiring recipients to personally evaluate the emails intended for them.
Thus, we designed a questionnaire to assess these emails and actively recruited participants for a comprehensive evaluation.

We categorized participants into two groups: academic researchers and non-academic professionals. In Section~\ref{sec:pii_collection}, we noted that information about academic researchers is publicly available online, facilitating easy collection by attackers. Consequently, we assumed that academic researchers would be more susceptible to targeted phishing attacks, prompting their inclusion in our study. 
We also included non-academic professionals to examine phishing susceptibility across diverse organizational contexts, expanding the analysis beyond academic settings.
We defined non-academic professionals as individuals employed within non-academic organizations.

We recruited academic researchers by posting experiment descriptions within university research communities.
Each community consists of people engaged in research including professors, graduate students, and undergraduate students. 
We recruited non-academic professionals through industry practitioners who have previously collaborated with the authors.
In total, we recruited 60 participants through Google Forms and categorized them based on the provided institutional email addresses to analyze response variations across institutional types: 55\% were academic researchers and 45\% were non-academic professionals.
All participants were informed about an incentive of approximately \$7.50 for their participation.

We emphasize that participants \textit{did not directly receive these emails}; instead, their responses were assessed through a Google survey.\footnote{The process with ethical/privacy-preserving measures is explained in Appendix~\ref{apx:recruiting_process}.} 
The rationale for adopting this survey format over a simulated phishing attack will be discussed in Section~\ref{sec:discuss}.
Each participant evaluated seven phishing emails generated using their own email as a target. The types of emails will be described in Email Design part.

\begin{table}[t]
\centering
\footnotesize
\caption{Types of LLM-crafted phishing emails.}
\label{tab:email_type}
\begin{tabular}{ c c c c }
\toprule
\textbf{Model} & \textbf{\makecell{Web\\search}} & \textbf{Purpose} & \textbf{\makecell{Sender's\\institution }} \\ 
\midrule
Claude & o & General & not designated \\
GPT& o & General & not designated  \\
Claude& x & General&  not designated \\
GPT & x & General & not designated \\
Claude& o & Credentials & not designated  \\
GPT & o & Credentials & not designated  \\
Claude & o & General & different from target's \\

\bottomrule
\end{tabular}
\end{table}

\noindent{\textbf{Email Design.}}
Using direct terms like ``phishing email''  would trigger the LLM’s safeguards and result in rejection.
Therefore, our experiment includes minimal circumvention techniques, such as avoiding such explicit terms.
Although these minimal efforts were sufficient for Claude and GPT, they could not bypass the safeguard restrictions of Gemini and refused generation.

As shown in Table~\ref{tab:email_type}, we designed seven distinct types of emails to evaluate the effectiveness across different variants of phishing emails.
We divided the purpose of email into \textit{general} and \textit{login credentials}. 
For the general purpose, the WebSearch agents were tasked to generate a realistic scenario in which the target might receive an email. 
To examine the effectiveness of web search functionality, we compared the emails generated by vanilla LLMs to those created by WebSearch agents.
For scenarios involving login credentials, we provided specific instructions for WebSearch agents to generate an email prompting the target to update their credentials via a designated link. 
Additionally, to examine the impact of the sender’s affiliation, we modified the prompt for the WebSearch agent using Claude to make the sender belong to a different institution than the target.

\noindent{\textbf{Questionnaire Design.}}
We designed seven questions to build an understanding of how participants perceive and interact with the emails, from surface content to deeper implications of authenticity and expected actions.
The first two questions focus on the \textit{content} of the emails, asking participants to identify the presented information (Q1) and to evaluate its accuracy against their actual personal details (Q2). 
Subsequent questions explore \textit{perceptions} of the sender and recipient, including whether the recipient identified the sender (Q3) and if the recipient's name correctly matched their real name (Q4).
We then inquire about the \textit{actions} participants would likely take upon receiving the email (Q5). 
The last question investigates \textit{perceived authenticity} of the email (Q6), asking participants to rate how genuine or fraudulent it appeared and to identify specific elements that influenced their judgment (Q7). 
The questionnaire is available at this link\footnote{\url{https://zenodo.org/doi/10.5281/zenodo.13691327}}.

\begin{figure}[t]
    \centering
\includegraphics[width=0.45\textwidth]
    {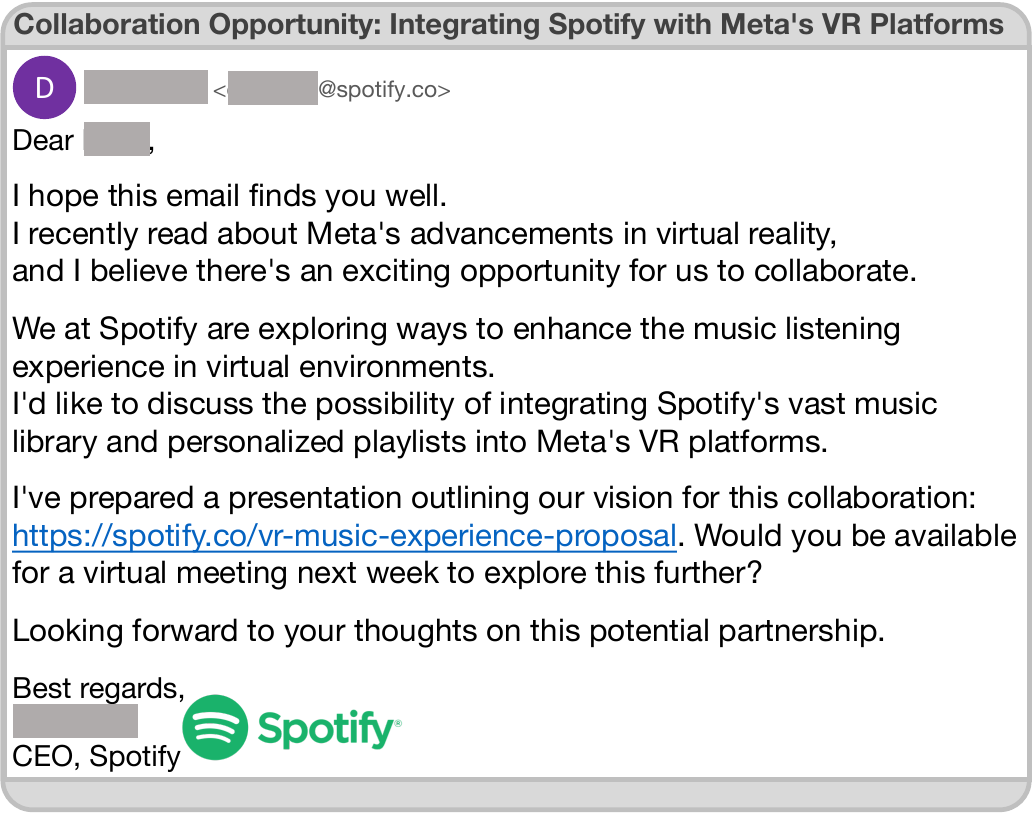}
    \caption{Example of a phishing email to a well-known figure generated by the WebSearch agent with Claude using an email address only.}
    \label{fig:phishing_demo}
    \vspace{-0.3cm}
\end{figure}

\subsection{Main Results}
We first analyze the effectiveness and content of \textit{general-purpose} emails created by vanilla LLMs and WebSearch agents. 
Then, we examine the results based on the participants' groups. 
We also analyze the efficacy of emails in relation to their intended purpose and the sender's institution.
In Tables~\ref{tab:overall} (Appendix~\ref{apx:phishing_eval_results}), we report the comprehensive survey results concerning participants' anticipated actions for various email types.

\begin{figure}[t]
    \centering
\includegraphics[width=0.42\textwidth]
    {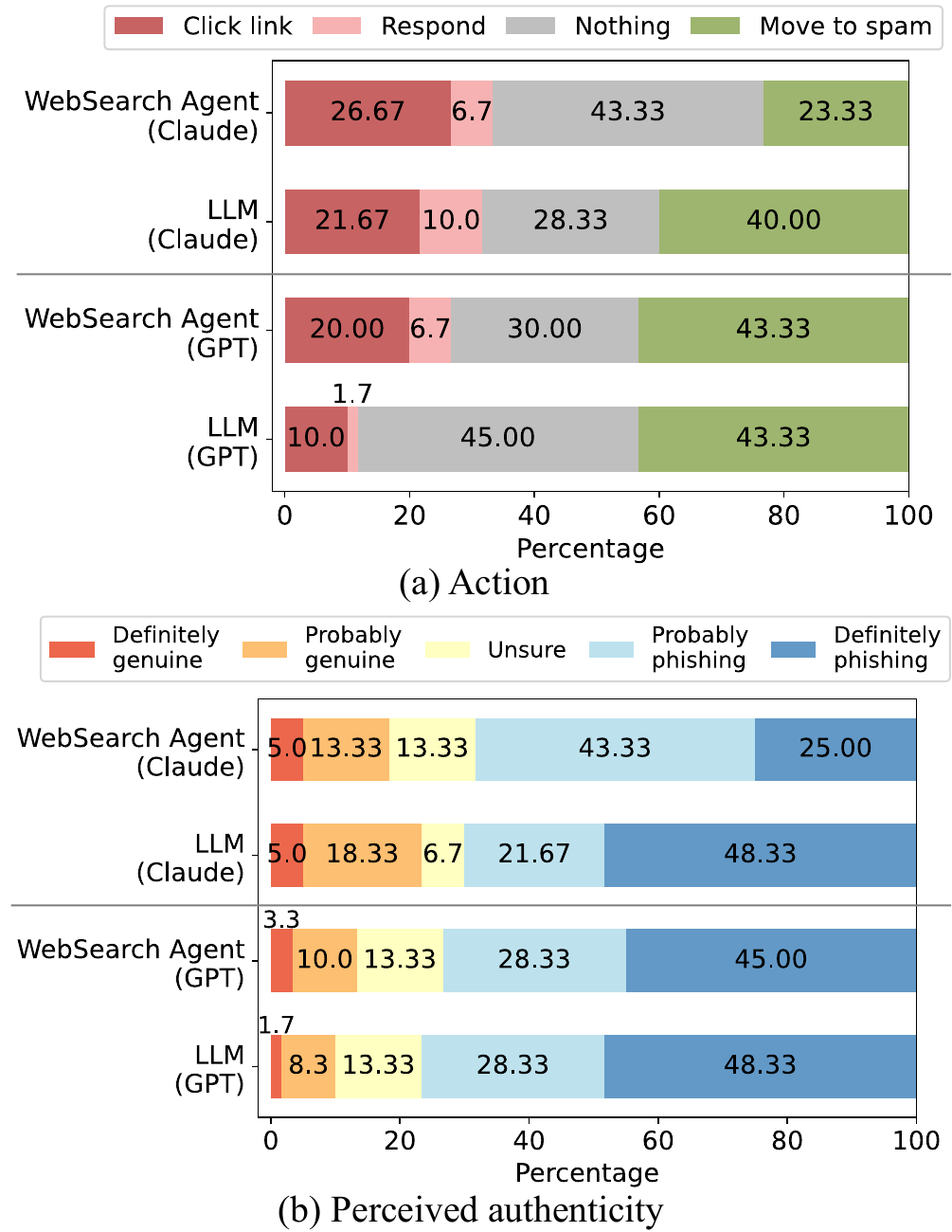}
    \caption{Survey results on participants' (a) expected actions upon receiving emails generated by LLMs and WebSearch agents, and their (b) perceived authenticity.}
    \label{fig:phishing_main_result}
    \vspace{-0.3cm}
\end{figure}

\subsubsection{Overall Analysis}
Figure~\ref{fig:phishing_demo} provides an example of phishing email generated by WebSearch agent using Claude.
We prompted the agent to write an email to a well-known figure, by providing only his publicly available email address.
The agent generated a plausible email considering his information, including a realistic-looking URL, and proposed a collaboration between Meta and Spotify.

\noindent{\textbf{Effectiveness Analysis.}}
Figure~\ref{fig:phishing_main_result} (a) presents the survey results of actions participants would likely take upon receiving the email (Q5).
Our primary focus was on the proportion of participants clicking on links, directly aligning with the attacker's goal.
The results indicate that WebSearch agents prompt more link clicks than vanilla LLMs, with the WebSearch agent using Claude achieving the highest click rate at 26.67\%.\footnote{Expert-crafted spear phishing emails in \cite{bethany2024large} show 26.6\% click rate. Detailed discussion in Section~\ref{sec:factor_phishing}.} 
In the case of GPT, the WebSearch agent's click rate was twice that of the vanilla LLM.
When comparing Claude and GPT, Claude was more effective, with its vanilla LLM variant achieving a slightly higher click rate than the WebSearch agent using GPT.
The disparities in effectiveness between Claude and GPT may be attributed to differences in email content, which will be explored further in the subsequent section.

When comparing ``actions'' (Q5) and ``perceived authenticity'' (Q6) in Figure~\ref{fig:phishing_main_result}, the percentage of participants identifying the email as \textit{definitely phishing} closely aligned with those opting \textit{move to spam}. 
\leftdiffnote{MR5}\diff{
On average, 66.27\% of respondents who identified an email as \textit{definitely phishing} opted for \textit{move to spam} as their action.}
However, the percentage of participants who rated the email as \textit{genuine} (definitely genuine + probably genuine) did not closely match the percentage who chose to \textit{click link}. 
Interestingly, for emails generated by WebSearch agents, the click rate was much higher than the rate of perceived genuineness, a pattern not observed with vanilla LLM-generated emails.
This discrepancy was driven by participants marked as \diffnote{MR5}\diff{\textit{unsure} or even \textit{phishing}; 13.34\% of participants stated that they would click on links in WebSearch agent-generated emails despite doubting their authenticity. In contrast, for vanilla LLM-generated emails, 3.34\% and 10\% did so for GPT and Claude, respectively.
Table~\ref{tab:contigency_table} (Appendix~\ref{apx:contigency_tab}) presents a contingency table showing the percentage distribution of participants’ actions (Q5) relative to their perception of authenticity (Q6).}
These results highlight that the WebSearch agents not only provided a sense of credibility but also effectively utilized target information from the web to engage the recipients' interest, leading to higher link-clicking rates.

\begin{figure}[t]
    \centering
    \includegraphics[width=0.45\textwidth]
    {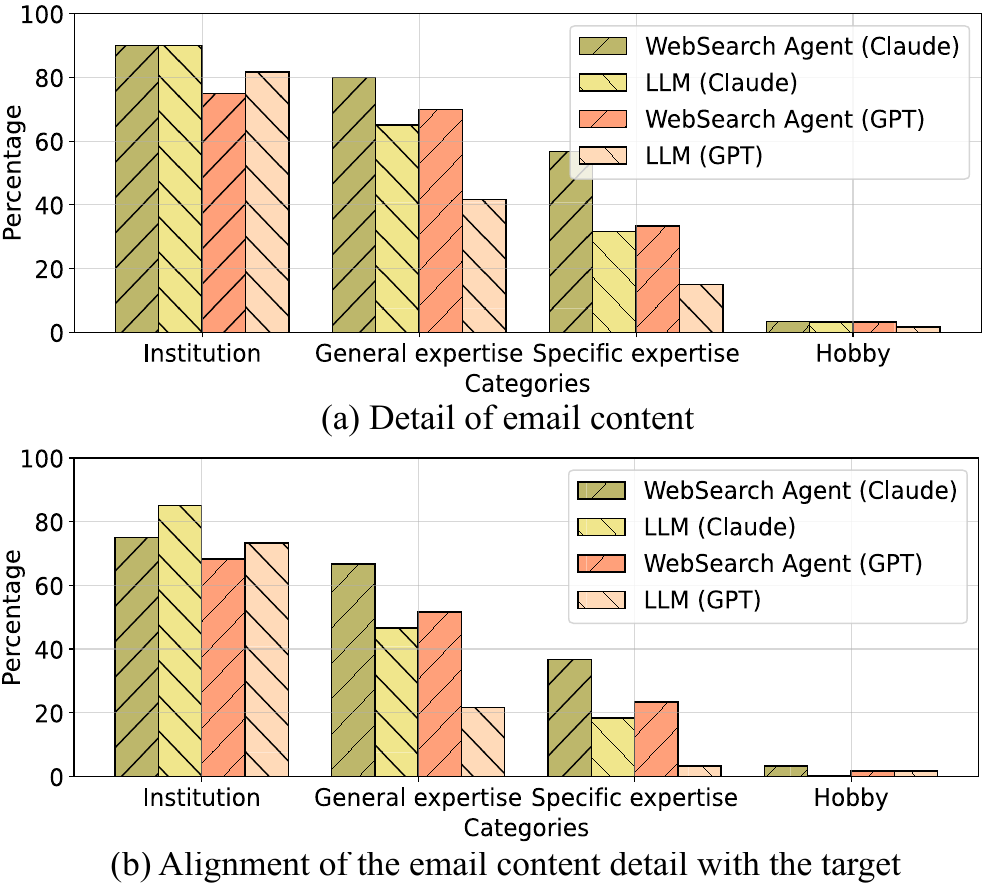}
    \caption{Analysis of email content: (a) detail of email content and (b) alignment of email content with targets. }
    \label{fig:phishing_content_result}
    \vspace{-0.3cm}
\end{figure}

\noindent{\textbf{Content Analysis.}}
\diff{We analyzed which email aspects influence participants' perceptions of authenticity using chi-squared tests.
\diffnote{MR5}From the analysis, the alignment of \textit{the details of email content} and the \textit{target's personal information} played statistically significant roles in participants' assessment of an email's authenticity. 
Table~\ref{tab:combined_chi_square} in Appendix~\ref{apx:chi_square} provides statistical results of the chi-squared tests.}

Building on these statistical results, we further examined the influence of web accessibility on the two factors.
Figure~\ref{fig:phishing_content_result} presents a comparative analysis of the percentage of phishing emails that incorporate specific content details and how well this content aligns with the targets' actual information, based on participants' evaluations.
When comparing GPT and Claude, it was noted that GPT-generated emails were less detailed, regardless of web usage.
We also observed that emails crafted by WebSearch agents were significantly richer in detail and more accurately reflected the target's personal data than those generated by vanilla LLMs.
This difference was particularly pronounced in the category of specific expertise. 
The results from our chi-squared tests in Table~\ref{tab:combined_chi_square} (Appendix~\ref{apx:chi_square}), indicate that the alignment of content with a target’s specific expertise significantly influenced the perceived authenticity of the emails.
By leveraging detailed web-sourced information about the target’s expertise, WebSearch agents were able to craft more convincing and effective phishing emails.

\noindent{\textbf{Factors Influencing Phishing Email Detection.}}
We examined the factors that influenced participants' perceptions of an email as \textit{phishing} (Q7, multiple responses allowed).
The factors were ranked by their influence, with sender information (62.3\%), email purpose (39.7\%), and information related to expertise (31.6\%) being the most significant. 
This result aligns with the previous studies showing that users heavily consider the sender's information when assessing an email's authenticity~\cite{10190709}. 
This finding suggests that employing spoofing techniques, such as manipulating the sender's email, could potentially lead to more effective attacks.


\subsubsection{Comparison Between Participant Groups}
\begin{figure}[t]
    \centering
    \includegraphics[width=0.42\textwidth]
    {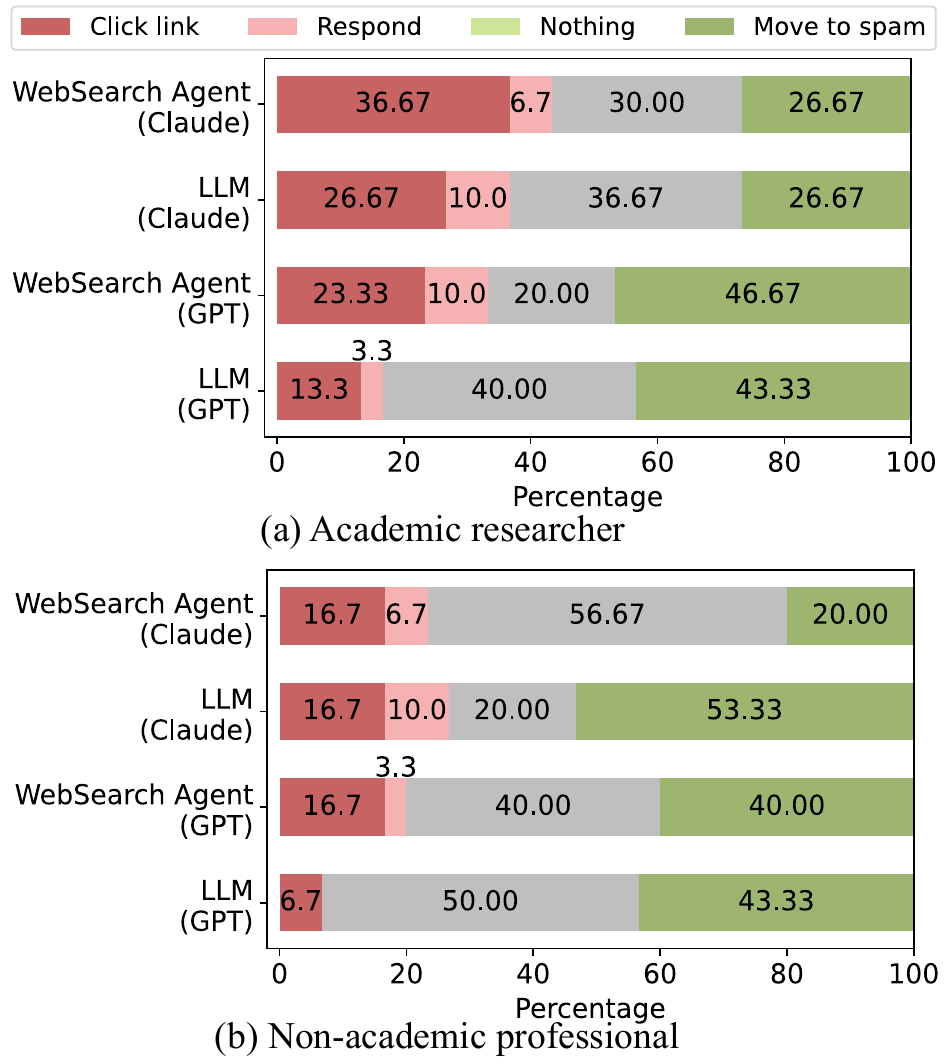}
    \caption{Analysis according to group: (a) academic researchers and (b) non-academic professionals. }
    \label{fig:phishing_group_result}
\end{figure}
Figure~\ref{fig:phishing_group_result} presents the actions participants from different groups reported they would take after receiving an email.
The academic researcher group generally showed a higher likelihood of clicking on links compared to the non-academic professional group.
This difference may be associated with the varying percentages of participants who verified that the phishing emails did not contain accurate information (Q2).
In the academic research group, the proportion of emails considered inaccurate varied from 0\% to 12.1\%, suggesting that the majority of emails contained accurate information.
In contrast, in the non-academic professional group, even for WebSearch agent-generated emails, up to 40.7\% of participants reported that none of the information matched their own.
This is most likely due to the amount of available information online, with research labs and universities often dedicating websites that list their members with information such as current projects, research interests, and recent publications. 

For the academic research group, the WebSearch agents for both Claude and GPT were more effective at inducing link clicks compared to the vanilla LLMs, with increases of 1.37 times and 1.75 times, respectively. 
In the other group, adding web accessibility to Claude resulted in the same link click rate and significantly reduced the ``move to spam'' percentage. 
For GPT, adding web accessibility increased the link click rate by 2.49 times.
These findings indicate that WebSearch agents can craft effective phishing emails with web accessibility. 
Researchers, who typically have a considerable amount of their information publicly accessible online, are especially vulnerable to such attacks.

\subsubsection{Factors Increasing Phishing Effectiveness}
\label{sec:factor_phishing}
We examined the impact of changing the email's purpose and sender's address on the effectiveness of phishing attacks.
Emails targeting login credentials proved more effective in prompting recipients to click the provided link, with a click rate of 33.33\% for Claude's WebSearch agent and 25\% for GPT's WebSearch agent (see Table~\ref{tab:overall} in Appendix~\ref{apx:phishing_eval_results}). 
This aligns with existing research~\cite{alkhalil2021phishing}, which suggested that urgent requests, such as login information updates, are effective in phishing scenarios.

Another notable observation is that using a sender's email address from an organization different from the target's resulted in the highest click rate, at 46.67\%. 
This rate is 1.75 times higher than that for the general purposed email generated by the WebSearch agent of Claude, as detailed in Table~\ref{tab:overall}.
Participants were less familiar with domains from other organizations, making them more susceptible to deception.

An interesting point is that our results show WebSearch agents could be as effective as, or even more effective than, human-crafted spear phishing emails in prompting clicks.
In the previous research~\cite{bethany2024large}, link click rates for spear phishing emails using internal email addresses and information were 26.6\% for human-crafted and 16.7\% for LLM-generated emails.
In our study, over 20\% of participants (up to 46.67\%) responded they would click on links in WebSearch agent-crafted emails. 
While these results are not directly comparable due to different methods and participant groups, they suggest that LLM agents were highly effective in generating phishing emails.
This is particularly noteworthy considering that our attack model requires significantly less knowledge and capability, and involves no human labor.

\section{Approachability of Cyberattacks}
\label{sec:approach}
In this section, we examine the practicality of using LLM agents for targeted cyberattacks by assessing two primary factors: cost and safeguard capability. 
Understanding these elements is crucial to determining the practicality of employing LLM agents for malicious purposes in real-world scenarios. 

\subsection{Cost Analysis}
\label{sec:cost}

Conducting attacks incurs both time and financial costs, which can highly impact the practicality of using LLM agents for cyberattacks. 
We examine how quickly and at what cost attackers can carry out cyberattacks using commercially available LLMs.
We measured the time spent and the number of tokens consumed during each attack scenario using \textit{WebSearch agent} for each individual target. 
To simplify our calculations, we considered token usage as the sum of input and output tokens, applying the output token rate, which is typically more expensive than the input token rate.
This approach ensures we estimate the upper limit of API costs.
For the models utilized, the costs are as follows: GPT 4o~\cite{openai_api_price} and Claude 1.5 Sonnet~\cite{anthropic_api_price} at \$15/1M tokens, and Gemini-1.5 Flash~\cite{gemini_api_price} at \$0.30/1M tokens.
To measure the number of tokens, we utilized the tiktoken tokenizer~\cite{tiktoken_openai} for GPT, as specified by OpenAI~\cite{tiktoken_openai}. 
For Claude and Gemini, we used the token counting feature provided by their APIs~\cite{anthropic_tokencnt, gemini_tokencnt}.

On average, collecting PII for an individual required 9.1 seconds and incurred a cost of 2.2 cents (GPT: 6.7s, 2.1\textcent; Claude: 11.4s, 2.2\textcent).
For generating impersonation posts, GPT, Claude, and Gemini took an average of 9.4, 21.7, and 3.6 seconds, respectively, with costs under 3 cents (GPT: 2.2\textcent; Claude: 3.0\textcent; Gemini: 0.03\textcent).
For creating phishing emails, GPT and Claude took an average of 7.1 and 22.7 seconds, respectively,  with costs below 4 cents (GPT: 2.7\textcent; Claude: 3.9\textcent).
See Appendix~\ref{apx:appendix_cost} for the cost statistics.

Cost variations across different LLMs can be attributed to differences in model capabilities, as analyzed in \cite{gpt-4o_analysis}. 
Considering all scenarios, WebSearch agents prove to be highly practical, executing tasks quickly and at minimal cost.
WebSearch agents can exploit private data in cyberattacks very quickly and at minimal cost, emphasizing their practicality for malicious use. 

\subsection{Safeguard Capability}
We conduct an in-depth analysis of safeguard capabilities by exploring different factors in impersonation post generation and spear phishing email generation.

\noindent{\textbf{Impersonation Post Generation.}}
In Section~\ref{sec:impersonation}, we found that the activation of safeguards varied depending on the nature of the claim. 
For further analysis of the safeguard capabilities of each model, we used two additional claims (``{Invest in Dogecoin}'' and ``{LLMs are effective in generating phishing emails}'') and assessed their safeguard bypassing when paired with different groups of targets.
These two groups were 10 professors from previous analyses and 10 influential figures (targets with more fame and general popularity) in the tech industry as targets for impersonation.

The activation of safeguards varied not only based on the nature of the claim but also on the nature of the impersonation target. 
In this task, we observed different safeguard capabilities in the services, with Claude being the most strict and GPT being the most tolerant of prompts.
For GPT, the vanilla LLM and WebSearch agent successfully generated posts for both new claims without any rejections.
However, Claude and Gemini showed different results.
Whereas WebSearch agents generated posts for the ``{Invest in Dogecoin}'' claim, their respective vanilla LLMs produced content for professors but rejected creating content for influential figures at rates of 60\% and 20\%.
When prompted with the claim about the harmful use of LLM, for Gemini, both the WebSearch agent and vanilla LLM successfully generated content, while Claude refused to generate content, responding, \textit{``I will not create content promoting harmful uses of AI systems''}.
The safeguard's dependence on the target is a characteristic that may restrict an attacker's range of targets.

\noindent{\textbf{Spear Phishing Email Generation.}}
The safeguard assessment in LLMs against spear phishing email generation involved varying two key factors: 1) the purpose of the email (general, request login credentials, request money), and 2) the type of email address used (institutional or Gmail). 
We generated emails using the email addresses of the authors, with their consent. 

Gemini shows the strongest safeguards by refusing to generate emails in all scenarios of vanilla LLM and WebSearch agent settings, responding with messages like \textit{``I cannot fulfill your request. Creating emails without consent is unethical and potentially illegal''}.
All models without web accessibility refuse generating content when the email's purpose involves requesting login credentials or money, indicating these topics trigger stricter safeguard mechanisms.
Conversely, general-purpose emails are less likely to be blocked, with GPT and Claude generating responses in these cases.
GPT and Claude are more likely to generate an email when provided with an institutional email address, compared to a Gmail address. 

Interestingly, when web accessibility was added, WebSearch agents, which previously refused to generate content under strict settings, began to produce responses.
Web-tools increased safeguard bypass of GPT and Claude, allowing generation in most of our experimental settings.
This demonstrates that web-based tools may introduce a vulnerability for jailbreaking LLMs.


\diff{
\section{Discussion and Limitations}
\label{sec:discuss}
\leftdiffnote{MR1}
\subsection{Adoption of Survey Format}
In our study, we chose a survey-based approach to understand user behavior towards phishing emails due to the following considerations.

\noindent{\textbf{Understanding Intentions and Reasoning.}} 
Survey-based research allows us to explore the reasoning and intentions behind participants’ actions, providing insights that are difficult to capture through mere analysis of click rates.
Such methods have been endorsed in prior research~\cite{10190709, huang2023evaluating, li2019analysis} as effective means of understanding user behavior in security settings.

\noindent{\textbf{Addressing Ethical Concerns.}} Simulating phishing attacks often involve deceptive practices that raise significant ethical concerns~\cite{schops2024simulated, busch2016ethical, phishing_ethic}.
Prior studies using simulations tend to be restricted to single organizations where comprehensive controls and institutional reviews are possible~\cite{lain2022phishing, bethany2024large}.
These controlled environments, while beneficial for managing ethical risks, restrict the diversity of the participant pool.
By opting for a survey-based approach, we were able to gather valuable data across 25 different organizations without encountering ethical risks.

\subsection{Defense}
We suggest defense strategies against the unauthorized collection of PII by LLMs utilizing web-based tools. First, \textbf{LLM service vendors} can implement a rule requiring LLMs that use web crawling tools to check and adhere to the directives specified in the \texttt{robots.txt} files of websites. 
These files serve as an international recommendation that either allows or restricts web crawlers from collecting site and webpage data. 
By respecting these directives, LLMs can prevent the inadvertent or intentional scraping of sensitive information from restricted areas of websites.
Additionally, scalable safeguards can be deployed, as further discussed in Section~\ref{sec:scalable safeguards}.

Second, PII providers (e.g., website managers) responsible for handling sensitive information should proactively control and reduce its online exposure.
A straightforward measure involves crafting and regularly updating robust \texttt{robots.txt} files that explicitly block crawlers from accessing pages containing personal information. 
In addition, they can employ more creative tactics to ``trick'' LLMs into gathering incorrect or incomplete PII. 
For instance, a website might display scrambled or false personal data to automated crawlers, while showing the correct information only after a user takes another intentional action. 
This approach ensures that human visitors see the accurate details, but an LLM performing automated scraping inadvertently collects erroneous information, thereby reducing the potential for unauthorized access or misuse.

\subsection{Risks with Expanding Capabilities}
\label{sec:scalable safeguards}
\leftdiffnote{MR7}The evolution of LLMs has not only linked their expanding capabilities to increased proficiency in executing complex tasks but also highlighted their potential misuse in scenarios such as cyberattacks.
Our findings demonstrate that as LLMs are enabled with additional tools, their effectiveness in cyberattacks is significantly enhanced. 
This correlation between enhanced capabilities and the associated risks underscores the necessity for proportionately stronger safeguards.

In response to these growing concerns, companies like Anthropic have developed scalable safeguards with the potential risks~\cite{anthropic_safegaurd_update}.
This approach involves setting specific thresholds at which the abilities of an AI system necessitate enhanced safeguards. 
Upon reaching these thresholds, the system automatically enforces stronger security measures tailored to the level of risk posed by the AI's capabilities. 
As the capabilities of LLMs increase, implementing scalable and adaptive security measures will be paramount to ensure their safe and ethical utilization.}

\subsection{Limitations}
In Section~\ref{sec:impersonation}, one of our primary objectives was to evaluate the effectiveness of LLM agents in generating impersonation posts according to their capability.
Understanding the link between tool usage and impersonation effectiveness required experiments to be conducted under uniform conditions, such as consistent target claims. 
However, collecting authentic posts from our targets—CS professors— regarding specific claims was infeasible. 
Thus, we were unable to use actual samples for comparison. 
Despite this, the inability to use authentic posts did not undermine our study's primary objective.

We also relied on LLM evaluators to judge the impersonation posts. Although these evaluators showed high agreement rates compared to human reviewers, they may not fully capture the nuanced reasoning and contextual awareness that human experts bring.

In Section~\ref{sec:phishing}, we focused on analyzing how participants interacted with LLM-generated phishing emails. 
While including synthetic non-phishing emails could provide a valuable baseline for comparison, this aspect was not incorporated into the current study.
We suggest it as a promising direction for future research to enhance our understanding of user responses to both genuine and phishing emails.

\section{Conclusion}
In this work, we examine the emerging threats posed by LLM agents, particularly their potential use of web-based tools in cyberattacks that exploit private data. 
Specifically, we focused on three types of cyberattacks: PII collection, impersonation post generation, and spear phishing email generation. 
Our experimental results demonstrate that attackers can successfully automate these attacks using LLM agents, with web-based tools enhancing the performance of these attacks.
Furthermore, our results report easy bypass of LLM safeguards at low-costs, making utilization of LLM agents in cyberattacks a viable option.
These findings expose a significant vulnerability in current safeguards and underscore the urgent need for security measures to prevent the misuse of LLM agents.

\section*{Acknowledgments}

This work was partly supported by Institute for Information \& communications Technology Technology Planning \& Evaluation(IITP) grant funded by the Korea government(MSIT) (RS-2023-00215700, Trustworthy Metaverse: blockchain-enabled convergence research, 50\%), (RS-2019-II190075, Artificial Intelligence Graduate School Support Program(KAIST), 20\%), and (RS-2024-00436680, Global Research Support Program in the Digital Field program, 20\%). It was also supported by Artificial intelligence industrial convergence cluster development project (10\%) funded by the Ministry of Science and ICT(MSIT, Korea)\&Gwangju Metropolitan City.

\section*{Ethical Considerations}
This paper discusses and examines the potential threat of LLM agents in cyberattacks exploiting personal data, which can raise ethical considerations. 
However, we minimize these concerns by employing the following measures:

\noindent\textbf{Aggregation of PIIs.}
For annotation purposes, these data were provided to annotators in Excel file format. Once the annotation process was completed, the annotators returned the annotated files to the principal investigator and deleted their local copies. After acceptance of our paper, all files containing personal information were deleted.

\noindent\textbf{User Study.} According to the local IRB's protocol, our research is excluded from the review for the following reasons. First, we used only publicly available information and did not handle any sensitive data. Additionally, the emails evaluated contained no harmful content, minimizing any potential impact on participants. In the recruitment process, we obtained informed consent from participants, detailing the purpose of the research, the expected duration, procedures involved, anticipated benefits, personal information protection, and consent for collecting and using their provided email addresses. 
We also informed participants that the LLM agents could search the internet for their email addresses to gather information and use it to generate emails. 
We asked them to evaluate emails through a Google survey after notification. Each participant was exposed only to emails generated from their own email address. 
\diff{All participant-related information was destroyed after the experiments, and participants were informed of this process. Additionally, they were assured that their responses would be excluded from the analysis if they withdrew consent, ensuring robust protection of their personal information.}

\diff{\noindent\textbf{Responsible Disclosure Process.} We plan to share our findings to the LLM service vendors. 
Specifically, we will share detailed prompts and raise concerns regarding the safeguard vulnerabilities of web-based tools.
Additionally, we will request the implementation of a rule ensuring that when an LLM utilizes web crawling tools, it checks and adheres to robots.txt directives.
Also, we strongly recommend that the PII providers from whom we collected data create a robots.txt file to prevent web crawlers from accessing pages containing personal information.}

\section*{Open Science Policy Compliance}
Our study explores the potential threat posed by LLM agents in cyberattacks. We acknowledge that disclosing the exact prompts used for these attacks could enable malicious actors to replicate them. To mitigate the risk of misuse, we have submitted our code for review with dummy prompts accessible at this link\footnote{\url{https://zenodo.org/doi/10.5281/zenodo.13691327}}.
We will provide the full source code only upon request for legitimate research purposes.




\bibliographystyle{plain}
\bibliography{reference}

\newpage
\appendix
\section{Details for Figure~\ref{fig:intro}}
\label{apx:intro}
In Figure~\ref{fig:intro}, the `LLM agent' denotes the highest capability setting for each task. 
For tasks involving PII collection and impersonation, the term refers to WebNav agents, whereas for spear-phishing email generation, it corresponds to WebSearch agents. 
The performance metrics vary according to each task and are denoted by the axes in the figure.

For the PII collection task, the metric represents the average number of PII items successfully collected from CS professors. 
In the case of the impersonation post generation task, it is measured by the percentage of posts judged as authentic in a binary (yes/no) authenticity test. 
For the spear-phishing email generation task, it indicates the click-through rate of links contained within emails designed for general purposes.

\section{PII collection - Human annotation}
\label{apx:pii_annotation}
To evaluate the accuracy of the collected PII, we employed human annotators to review the responses generated by LLM agents.
Six annotators were employed, all of whom held at least a bachelor's degree and were proficient in English.

Due to the vast and diverse nature of the field of CS, we did not evaluate based solely on department listings.
Instead, annotators were directed to classify individuals as CS professors based on their engagement in relevant fields, following a thorough web search.
Only for individuals identified as CS professors, annotators conducted the remaining four PII verification processes.
Certain types of PII, such as phone number, are prone to changes over time, complicating the verification process due to discrepancies across different web sources. 
Furthermore, the probabilistic nature of responses from LLMs can lead to inconsistent information retrieval. 
To address these concerns, we initiated five separate queries to LLMs for each individual. 
Annotators were then instructed to consider the information collection successful if at least one of the five responses matched the data available online.

During the annotation process for the names of CS students, annotators referenced the student information provided at the professors' laboratory introduction pages as the ground truth. 
The methodology for annotating other types of PII remained consistent with the approach used for professors' PII.

\section{Impersonation Post Evaluation Templates}
\label{apx:impersonation_eval_prompt}
\diffnote{RC5}\diff{
In Section~\ref{sec:impersonation}, LLM evaluators were utilized in both the A/B Test and the Yes/No Test. The prompts provided are detailed below.
These prompts were designed to enable LLMs to incorporate personal information, similar to how human evaluators use factual details like occupation and interests to assess post authenticity. 
This approach is particularly relevant in the context of advanced impersonation attacks, which exploit personal data to deceive effectively.
Experimentally, we observed that without such prompts, LLMs tend to rely on fluency, favoring polished entries while failing to evaluate authenticity. 
By guiding models, we could align their evaluation process more closely with human judgment.
}
\begin{tcolorbox}[colback=white, colframe=black, title=A/B Testing Prompt]
\label{fig/prompt1}
\footnotesize
This is a social media post (like Twitter or Facebook).
Decide which post appears more likely to have been created by the author themselves, rather than generated by AI or another person. 
You can use accurate personal information of the author from a web search.
Respond with either [answer]postA[answer] or [answer]postB[answer] or [answer]unsure[answer]. PostA: \textbf{TextA}, PostB: \textbf{TextB}
\end{tcolorbox}

\begin{tcolorbox}[colback=white, colframe=black, title=Yes/No Testing Prompt]
\label{fig/prompt2}
\footnotesize
Do you think this content was written by the author themselves, rather than being generated by AI or another person? 
You can use personal information of the author from a web search.        
Respond with either [answer]Yes[answer] or [answer]No[answer]. \textbf{Text}.
\end{tcolorbox}

\section{Approachability of Cyberattacks}
\label{apx:appendix_cost}

Table~\ref{tab:cost_stat2} represents a cost analysis of PII collection, impersonation post generation, and phishing email generation using WebSearch agents.

\begin{table}[h]
\centering
\caption{Cost statistics for WebSearch agents to generate impersonation post and spear phishing email.}
\label{tab:cost_stat2}
\footnotesize
\setlength{\tabcolsep}{4pt}
\begin{tabular}{ccc rrr}
\toprule
&&&\textbf{GPT} & \textbf{Claude} & \textbf{Gemini} \\ 
\midrule
\multirow{6}{*}{\textbf{PII Collection}} & \multirow{3}{*}{Time} 
& min &  3.3 & 7.9  & N/A\\
&& max & 11.2 &  7.51 & N/A \\
&& mean & 6.7 & 11.4 & N/A\\
\cmidrule(r){2-6}
& \multirow{3}{*}{\# Tokens} 
& min &  438.0 & 1159.0 & N/A\\
&& max & 2559.0 & 2268.0 & N/A\\
&& mean & 1406.0 & 1462.8 & N/A\\
\midrule
\multirow{6}{*}{\textbf{Impersonation}} & \multirow{3}{*}{Time} 
& min &  2.0 & 6.5 & 2.7\\
&& max &  24.2 & 33.0 & 4.6\\
&& mean & 9.4 & 21.7 & 3.6 \\
\cmidrule(r){2-6}
& \multirow{3}{*}{\# Tokens} 
& min &  519 & 1122 & 772 \\
&& max &  2640 & 2751 & 1075\\
&& mean & 1435.1 & 1993.2 & 917.7 \\
\midrule
\multirow{6}{*}{\textbf{Phishing}} & \multirow{3}{*}{Time} 
& min &  4.9 & 18.0 & N/A\\
&& max & 11.1 &  43.9 & N/A \\
&& mean & 7.1 & 22.7 & N/A\\
\cmidrule(r){2-6}
& \multirow{3}{*}{\# Tokens} 
& min &  803.0 & 1984.0 & N/A\\
&& max & 2760.0 & 3761.0 & N/A\\
&& mean & 1803.7 & 2623.0 & N/A\\
\bottomrule
\end{tabular}
\end{table}

\section{Spear Phishing Email Generation}

\subsection{User Study}
\label{apx:recruiting_process}
At the beginning of the study, participants were provided with a consent form. 
Then, we asked to provide their institutional email addresses.
Upon receiving consent, we collected email addresses for the purpose of data collection integral to the experimental design. 
For each email address, we generated seven emails using LLMs and WebSearch agents.

Participants were then asked to assess these emails through a questionnaire on Google Survey that we administered. 
It is important to note that each participant was exposed only to emails generated from their own email address.
We instructed participants to evaluate the emails under assumption that they had received these emails to their own email addresses.

\diff{\subsection{Chi-Squared Tests}
\label{apx:chi_square}
\leftdiffnote{MR5}This section investigates the factors influencing participants' perception of email authenticity.
First, we identify which aspects of an email had the most significant influence on determining the email's authenticity.
We conducted chi-squared tests to evaluate the independence between email evaluation questions (Q1–Q4) and the question assessing perceived email authenticity (Q6). 
For details on the questions used in our survey, please refer to the document titled `Questionnaire for User Study,' accessible at this link\footnote{\url{https://zenodo.org/doi/10.5281/zenodo.13691327}}.
In Table~\ref{tab:combined_chi_square}, the results for ``content detail'' and ``content authenticity'' show p-values significantly less than 0.05, indicating strong evidence against the null hypothesis.

Further analysis focused on ``content authenticity'' (Q2), which was identified as the most strongly associated factor. 
Specifically, we examined which aspects of email content influenced participants' perceptions of authenticity. 
We evaluated the independence between each response option within Q2 and the option in Q6 using chi-squared tests. 
The ``Specific area of expertise'' option had a p-value much less than 0.05, indicating a highly significant association.
}

\subsection{Overall Survey Results}
\label{apx:phishing_eval_results}
Table~\ref{tab:overall} represents the survey results about participants' actions (Q5) across seven email types.

\diff{\subsection{Contingency Table}
\label{apx:contigency_tab}
\diffnote{MR5}Table~\ref{tab:contigency_table} presents contingency tables illustrating the percentage distribution of participants' actions (Q5) in relation to their perception of authenticity (Q6). The analysis focuses on emails with general purposes and no specified sender institution.
}

\begin{table*}[h]
    \centering
    \caption{Chi-squared test results: Comparisons of Q1-Q4 and Q6, and response options within Q2 and Q6.}
    \label{tab:combined_chi_square}
    \footnotesize
    \begin{tabular}{lcccc|cccc}
    \toprule
    \multirow{2}{*}{} &
    \multicolumn{4}{c|}{Questions} &
    \multicolumn{4}{c}{Response Options within Q2} \\
    \cmidrule(lr){2-5} \cmidrule(lr){6-9}
    & Q1 & Q2 & Q3 & Q4 & institution & \makecell{general expertise} & \makecell{specific expertise} & hobby \\
    &(content detail) & (content auth) & (sender) & (recipient) & & & & \\ 
    \midrule
    kappa& 98.80 & 113.54 & 56.54 & 3.69 & 4.34 & 0.01 & 11.15 & 0.28 \\
    p-value& $2.23 \times 10^{-5}$ & $3.12 \times 10^{-7}$ & 0.0159 & 0.4497 & 0.0595 & 0.0373 & 0.9436 & 0.0008\\
    \bottomrule
    \end{tabular}
\end{table*}

\begin{table*}[t]

\centering
\footnotesize
\caption{Survey results about expected actions upon receiving emails. Percentages represent responses relative to the total number of participants.}
\label{tab:overall}
\begin{tabular}{@{}lccccccc@{}}
\toprule
 & &&&  \multicolumn{4}{c}{Action} \\
\cmidrule(r){5-8}
    Model & Web & Purpose & Sender's institution  
     & Click link & Reply & Nothing & Move to spam\\
\midrule
\multirow{4}{*}{Claude} 

& o & general & not designated
& 26.67&   6.67  & 43.33 & 23.33\\
& x & general & not designated
 & 21.67& 10.00 & 28.33 & 40.00 \\
& o & credential &  not designated 
 & 33.33& 6.67 &25.00 & 35.00\\
  & o & general & different from target's 
& 46.67 &  3.33  &  41.67 & 8.33       \\
\midrule
\multirow{3}{*}{GPT} 
& o & general & not designated
& 20.00 &6.67  & 30.00 & 43.33\\
& x & general & not designated
& 10.00 & 1.67 & 45.00 & 43.33\\
& o & credential &  not designated 
 & 25.00  & 5.00 & 23.33 & 46.67\\

\midrule
\multicolumn{4}{c}{Average} & 26.19 & 5.71 & 33.81 & 34.29\\
\bottomrule
\end{tabular}
\end{table*}

\begin{table*}[h]
\footnotesize
\ifdiff
\leftdiffnote{MR5}
\begin{mdframed}[backgroundcolor=blue!20]
\fi
\centering
\caption{Contingency table showing the percentage distribution of participants' actions (Q5) against their perception of authenticity (Q6).}
\label{tab:contigency_table}
\begin{tabular}{clcccccc}
\toprule
&&  \multicolumn{5}{c}{Genuine or Phishing}  \\
\cmidrule(r){3-7}
\multicolumn{2}{c}{WebSearch Agent (Claude)}
&Definitely genuine
&Probably genuine
& Unsure
&Probably phishing
&Definitely phishing
& Total\\
\midrule
\multirow{4}{*}{Action} 
&Click link&5.00 &8.33 &1.67 &10.00 &1.67 &26.67 \\
&Reply&0.00 &0.00 &3.33 &3.33 &0.00 &6.67 \\
&Nothing&0.00 &5.00 &6.67 &21.67 &10.00 &43.33 \\
&Move to spam&0.00 &0.00 &1.67 &8.33 &13.33 &23.33 \\
\midrule
&Total&5.00 &13.33 &   13.33 &43.33 &25.00 &100.00 \\
\bottomrule
\toprule
&&  \multicolumn{5}{c}{Genuine or Phishing}  \\
\cmidrule(r){3-7}
\multicolumn{2}{c}{\makecell{LLM (Claude)}}
&Definitely genuine
&Probably genuine
& Unsure
&Probably phishing
&Definitely phishing
& Total\\
\midrule
\multirow{4}{*}{Action} 
&Click link&0.00 &11.67 &1.67 &5.00 &3.33 &21.67 \\
&Reply&5.00 &3.33 &0.00 &1.67 &0.00 &10.00 \\
&Nothing&0.00 &3.33 &5.00 &10.00 &10.00 &28.33 \\
&Move to spam&0.00 &0.00 &0.00 &5.00 &35.00 &40.00 \\
\midrule
&Total&5.00 &18.33 &6.67 &21.67 &48.33 &100.00 \\
\bottomrule
\toprule
&&  \multicolumn{5}{c}{Genuine or Phishing}  \\
\cmidrule(r){3-7}
\multicolumn{2}{c}{\makecell{WebSearch Agent (GPT)}}
&Definitely genuine
&Probably genuine
& Unsure
&Probably phishing
&Definitely phishing
& Total\\
\midrule
\multirow{4}{*}{Action} 
&Click link&1.67 &5.00 &6.67 &5.00 &1.67 &20.00 \\
&Reply&1.67 &1.67 &0.00 &3.33 &0.00 &6.67 \\
&Nothing&0.00 &3.33 &5.00 &10.00 &11.67 &30.00 \\
&Move to spam&0.00 &0.00 &1.67 &10.00 &31.67 &43.33 \\
\midrule
&Total&3.33 &10.00 &13.33 &28.33 &45.00 &100.00 \\
\bottomrule
\toprule
&\multicolumn{5}{c}{Genuine or Phishing}  \\
\cmidrule(r){3-7}
\multicolumn{2}{c}{\makecell{LLM (GPT)}}
&Definitely genuine
&Probably genuine
& Unsure
&Probably phishing
&Definitely phishing
& Total\\
\midrule
\multirow{4}{*}{Action} 
&Click link&1.67 &5.00 &1.67 &1.67 &0.00 &10.00 \\
&Reply&0.00 &0.00 &0.00 &1.67 &0.00 &1.67 \\
&Nothing&0.00 &3.33 &10.00 &16.67 &15.00 &45.00 \\
&Move to spam&0.00 &0.00 &1.67 &8.33 &33.33 &43.33 \\
\midrule
&Total&1.67 &8.33 &13.33 &28.33 &48.33 &100.00 \\
\bottomrule
\end{tabular}
\ifdiff
\end{mdframed}
\fi
\end{table*}


\end{document}